\documentclass[journal=jacsat,manuscript=article]{achemso}

\usepackage[version=3]{mhchem} 
\usepackage{xcolor}
\usepackage{amsmath}
\usepackage{siunitx}
\usepackage{color,soul}
\usepackage{overpic}
\usepackage{float}
\usepackage{tablefootnote}
\usepackage{threeparttable}
\usepackage{ulem}
\usepackage{soul}
\usepackage{hyperref}
\hypersetup{
  colorlinks=false,   
  pdfborder={0 0 0}    
}

\usepackage{xr} 

\newcommand{\beginsupplement}{%
        \setcounter{table}{0}
        \renewcommand{\thetable}{S\arabic{table}}%
        \setcounter{figure}{0}
        \renewcommand{\thefigure}{S\arabic{figure}}%
        \setcounter{equation}{0}
        \renewcommand{\theequation}{S\arabic{equation}}%
        \setcounter{section}{0}
        \renewcommand{\thesection}{S\arabic{section}}%
        \setcounter{subsection}{0} 
        \renewcommand{\thesubsection}{\thesection.\arabic{subsection}}%
        \setcounter{subsubsection}{0} 
        \renewcommand{\thesubsubsection}{\thesubsection.\arabic{subsubsection}}%
}

\author{Marta Monti}
\affiliation[ICTP]{Condensed Matter and Statistical Physics (CMSP), The Abdus Salam Centre for Theoretical Physics, Trieste 34151, Italy}
\email{mmonti@ictp.it}
\author{Yu Jin}
\affiliation[UCHI1]{Pritzker School of Molecular Engineering, The University of Chicago, Chicago, Illinois 60637, United States}
\author{Gonzalo Díaz Mirón}
\affiliation[ICTP]{Condensed Matter and Statistical Physics (CMSP), The Abdus Salam Centre for Theoretical Physics, Trieste 34151, Italy}
\author{Arpan Kundu}
\affiliation[UCHI1]{Pritzker School of Molecular Engineering, The University of Chicago, Chicago, Illinois 60637, United States}
\author{Marco Govoni}
\affiliation[UMORE]{Department of Physics, Computer Science, and Mathematics, University of Modena and Reggio Emilia, Modena 41125, Italy}
\alsoaffiliation[UCHI1]{Pritzker School of Molecular Engineering, The University of Chicago, Chicago, Illinois 60637, United States}
\alsoaffiliation[ARG]{Materials Science Division, Argonne National Laboratory, Lemont, Illinois 60439, United States}
\author{Giulia Galli}
\affiliation[UCHI1]{Pritzker School of Molecular Engineering, The University of Chicago, Chicago, Illinois 60637, United States}
\alsoaffiliation[ARG]{Materials Science Division, Argonne National Laboratory, Lemont, Illinois 60439, United States}
\alsoaffiliation[UCHI2]{Department of Chemistry, The University of Chicago, Chicago, Illinois 60637, United States}
\email{gagalli@uchicago.edu}
\author{Ali Hassanali}
\affiliation[ICTP]{Condensed Matter and Statistical Physics (CMSP), The Abdus Salam Centre for Theoretical Physics, Trieste 34151, Italy}
\email{ahassana@ictp.it}

\title[An \textsf{achemso} demo]
 {Defects at play: Shaping the photophysics and photochemistry of ice}

\begin{document}

\begin{abstract}
The mechanisms by which light interacts with ice and the impact of photo-induced reactions are central to our understanding of environmental, atmospheric and astrophysical processes. However, a microscopic description of the photoproducts originating from UV absorption and emission processes has remained elusive. Here we explore the photochemistry of ice using time-dependent hybrid density functional theory on various models of pristine and defective ice Ih. Our investigation of the excited state potential energy surface of the crystal shows that UV absorption can lead to the formation of hydronium ions, hydroxyl radicals and excess electrons. One of the dominant mechanisms of decay from the excited to the ground-state involves the recombination of the electron with the hydroxyl radical yielding hydronium-hydroxide ion-pairs. We find that the details of this charge recombination process sensitively depend on the presence of defects in the lattice, such as vacancies and pre-existing photoproducts. We also observe the formation of Bjerrum defects following UV absorption; we suggest that, together with hydroxide anions, they are likely responsible for prominent features experimentally detected in long UV exposure absorption spectra, remarkably red-shifted relative to short exposure spectra. Our results highlight the key role of defects in determining the onset of absorption and emission processes in ice.
\end{abstract}

The photochemistry and photophysics of ice play a crucial and complex role in diverse environments, including the interstellar medium and the Earth’s polar regions \cite{Grannas_2007,Crouse_2015}. For example, the icy surface present in polar regions traps a variety of volatile organic and inorganic compounds that, upon UV irradiation, are chemically transformed into photoproducts, strongly influencing the regions' atmospheric chemistry \cite{MACKAY199525}. Similarly, in the interstellar medium, the ice mantles formed on dust grains are thought to serve as sites for UV photochemical reactions \cite{Fillion_2022}. Hence, understanding how UV absorption affects the properties of water and ice has been an active area of research for decades, both experimentally and theoretically \cite{YABUSHITA_rev,Cuppen_2024}. 

UV absorption spectra of crystalline and amorphous ices, as well as of liquid water have been measured by several authors \cite{Onaka_1968,Kerr_1972,Heller_1974,Shibaguchi_1977,Minton_1971,kobayashi1983optical}. A common feature of these spectra is the presence of a peak in the range of 8.2-8.7 eV \cite{Kerr_1972,kobayashi1983optical}, preceded by a lower energy tail extending to 6-7.6 eV \cite{Williams_1976,Quickenden_1980,kobayashi1983optical}. Interestingly, the absorption of UV photons can trigger a complex set of chemical reactions in the excited state, eventually leading to the creation of radicals such as hydroxyl, in addition to other chemical species including hydrogen peroxide, oxygen and hydrogen gas \cite{matich1993oxygen,YABUSHITA_rev}. 

Quickenden and co-workers showed that UV-irradiation of both crystalline and amorphous ice for several hours leads to two pronounced absorption peaks at 220 nm (5.64 eV) and 260 nm (4.77 eV) \cite{quickenden1985uv,Quickenden1996,matich1993oxygen,langford2000}, much lower in energy than the peak and tail observed after a short time exposure to UV light. Emission following absorption at 5.64 eV gives rise to a short-lived fluorescence band at 340 nm (3.65 eV), while emission following absorption at 4.77 eV leads to a longer lived phosphorescence band at 420 nm (2.95 eV). Varied candidate photoproducts have been proposed as the source of fluorescence emission by Quickenden \textit{et al.} \cite{quickenden1985uv,matich1993oxygen,langford2000}, including OH radicals and molecular O$_2$.

However, the mechanism by which photoproducts are generated and their possible role in photoluminescence remain open questions. In addition, UV irradiation of ice studied in planetary science has been suggested to result in phase transitions, due to the creation of specific defects in the hydrogen bonded network \cite{Kouchi_2021}. To date, the precise origin of the electronic and molecular character of the excited state potential energy landscape of ice, the pathways that may give rise to the emission of photons in a broad energy range, and the role of different types of defects, remain largely unexplored.

In this contribution, we take an important step to fill this knowledge gap by deploying state-of-the-art excited state electronic structure calculations on various models of pristine and defective crystalline ice Ih. Using time-dependent hybrid density functional theory calculations, we explore the potential energy surface of the excited state of ice, and we determine the role played by different types of defects, including point vacancies, ionic charges and Bjerrum defects \cite{deKoning2020}, in near-to-mid UV absorption and emission processes. Our findings will help guide future experiments in the identification of photoproducts and they pave the way to building realistic models of the photophysics and chemistry of water in different thermodynamic conditions.

\section*{Results}

\begin{figure}
\centering
\includegraphics[width=\linewidth]{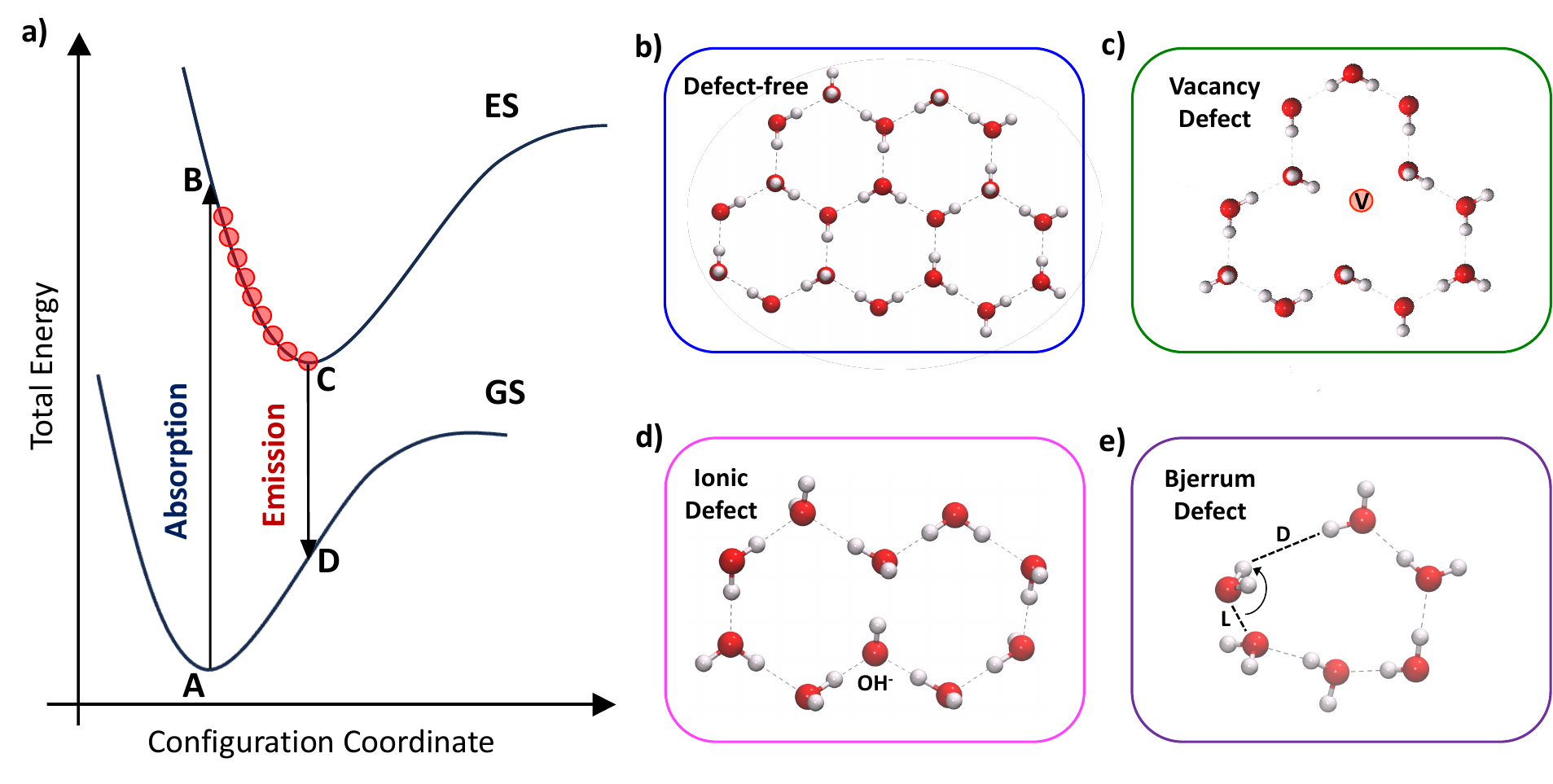}
\caption{Schematic representation of absorption and emission processes investigated in our work (a) for four different ice models: defect-free proton disordered ice (b); ice with vacancy defects (c), with ionic (OH$^-$) defects (d), and with Bjerrum defects (e). The ground and excited state potential energy surfaces are denoted by GS and ES, respectively. Hydrogen and oxygen atoms are represented by white and red spheres, respectively.}
\label{Fig1}
\end{figure}

Our strategy to investigate the interaction of UV light with ice leverages recent theoretical developments \cite{YuJin2023} that enable the efficient calculation of forces on atoms in the excited states (ES) for periodic systems, using time-dependent density functional theory (TDDFT) and hybrid functionals. This approach circumvents the need of using isolated clusters, previously adopted to study the photochemistry of water \cite{Chipman_2005,Chipman_2006,Sobolewski_2007,Suwannakham_2018}. A crucial ingredient to study ES potential energy surfaces (PES) is an appropriate exchange-correlation functional. Here, we use the dielectric-dependent hybrid (DDH) \cite{DDH_r1,DDH_r2} functional, which has been shown to provide an accurate description of the electronic properties of liquid and solid water \cite{Gaiduk2018,Bischoff2021} (see details in the \textit{Materials and Methods} section). 

To investigate the mechanism of photoluminescence in crystalline ice Ih (see Fig. \ref{Fig1}a), we first studied absorption and emission in a defect-free proton-disordered ice model (Fig. \ref{Fig1}b). We then considered various types of defects in the lattice (shown in panels c-e of Fig. \ref{Fig1}), to assess their impact on the optical properties of the solid, including vacancies, ionic (OH$^-$), and Bjerrum (L, D) defects. We considered absorption of a photon from configuration A to B, as shown in Fig. \ref{Fig1}a, followed by relaxation of the system geometry to the excited state configuration C and then emission of light from the ES to the D configuration on the ground state (GS) PES. In the following, we characterize the nuclear and electronic configurations A, B, C and D for pristine and defective ice.

\subsection{Defect-free Ice Ih}
\label{DF}

We begin by examining the optical properties of defect-free (DF), disordered ice Ih. We computed the onset of absorption for 100 lattice configurations, including the lowest 100 excitation energies for each configuration (see Fig. S1). We find consistent features for all configurations with an average onset energy of $\sim$9.5 ($\pm$ 0.0058) eV, as indicated by the red dashed line in Fig. S1. Here, the onset energies are defined as the lowest TDDFT excitation energies with non-zero oscillator strength, that is, the first allowed electronic transitions. This definition remains consistent across all structural models introduced in the following sections. 

\begin{figure}
\centering
\includegraphics[width=\linewidth]{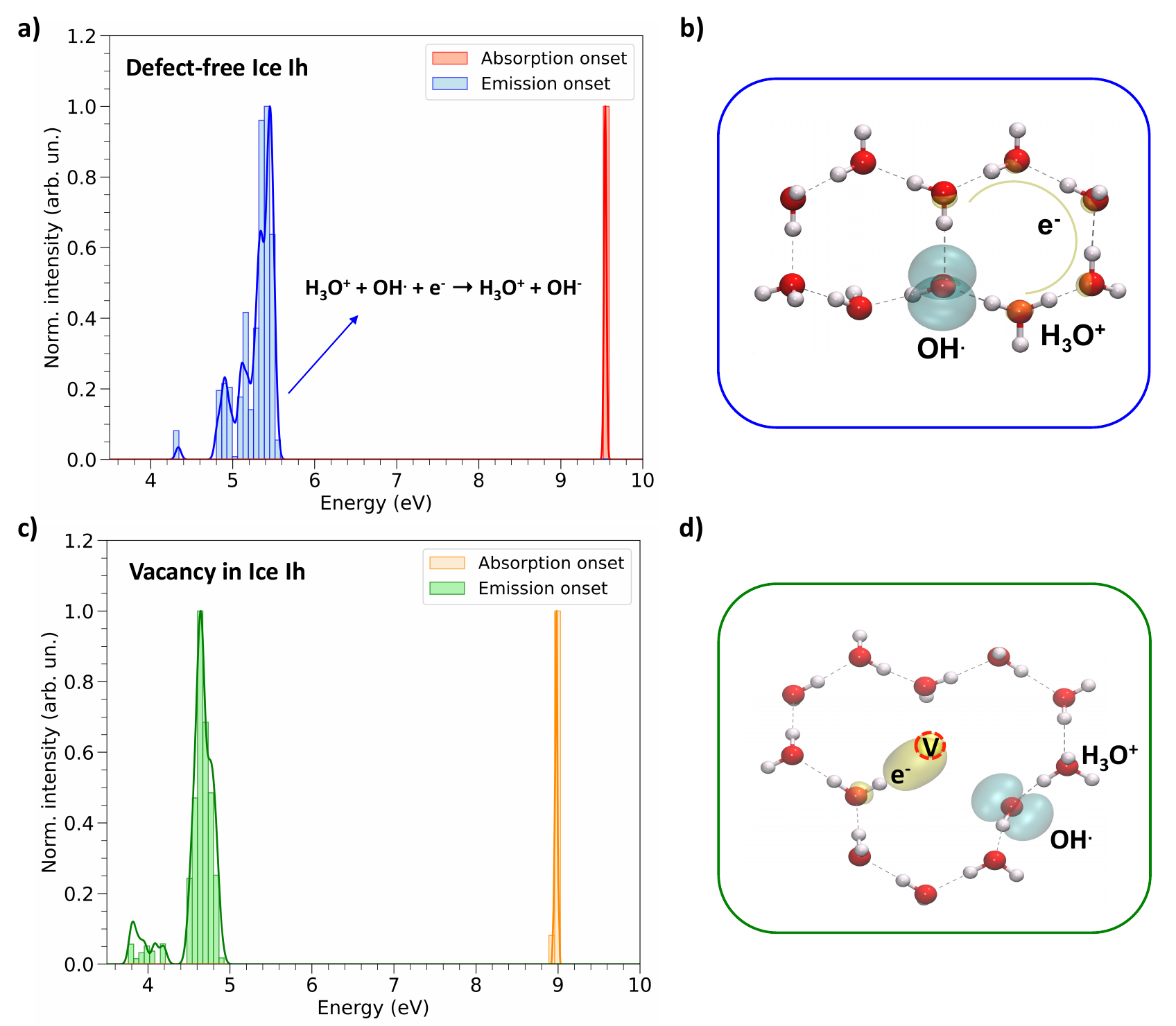}
\caption{Distributions of absorption (blue) and emission (red) onset energies for 100 configurations of defect-free, disordered ice Ih are shown in panel a. Solid lines represent Gaussian kernel density estimation (KDE) curves fitted to the corresponding normalized histograms (see \textit{Methods} section). Panel b shows the unrelaxed differential density computed for one representative configuration optimized in the excited state. The electron depletion and electron accumulation regions are displayed in blue and yellow, respectively. Distributions of absorption (orange) and emission (green) onset energies for 100 configurations of vacancy-containing, disordered ice Ih are shown in panel c, along with their respective Gaussian KDE curves. The unrelaxed differential density for one representative configuration is reported in panel d, with the same color scheme adopted in (b).}
\label{Fig2}
\end{figure}

The distributions of the onset absorption (A $\rightarrow$ B) and emission (C $\rightarrow$ D) energies are shown in Fig. \ref{Fig2}a as red and blue normalized histograms, respectively. The corresponding Gaussian kernel density estimation (KDE) curves are overlaid on the histograms. Interestingly, all configurations of DF show similar absorption energies, while the emission energies span a broad range of $\sim1.2$ eV. The lowest absorption energy corresponds to the transition between the valence band maximum (VBM) and conduction band minimum (CBM) of the solid. The occupied band edge is localized on specific oxygen atoms in the lattice, due to disorder in the crystal, and the CBM corresponds to a delocalized state, as shown in Fig. S2a. These findings are in agreement with those reported in several previous studies, including most recently by Tang \textit{et al.} \cite{Tang_2025}

Our computed absorption onset energy (9.5 eV) is larger than the experimentally reported 7.7 eV \cite{kobayashi1983optical}. Several studies over the last decade have shown that optical properties predicted from theory can be affected by a combination of the quality of the electronic structure, finite temperature and the inclusion of nuclear quantum effects (NQE). For example, previous studies solving the Bethe–Salpeter equation (BSE) and using the GW approximation have reported the absorption onset between 8 and 9 eV \cite{Hahn_2005,Hermann_2008,Nguyen_2019,Tang_2025}, depending on the level of theory used to obtain the single particle wavefunctions in the GW calculations. NQEs have been shown to substantially lower the fundamental energy gap of water and ice by 0.5-0.7 and 1.2-1.5 eV, respectively \cite{Engel_2015,Bischoff2021,Berrens_2024}. Recently, Tang \textit{et al.} \cite{Tang_2025} reported onset absorption energies blue-shifted by 0.8 eV, relative to experiments \cite{kobayashi1983optical}, in the absence of NQEs. Thus, the inclusion of the NQE corrections on the absorption edge of ice would move our onset energies into closer agreement with the experiments. As we will see later, besides these effects, the onset absorption energies are also sensitive to the presence of defects in the ice samples probed experimentally. Thus, a quantitative comparison should involve an understanding of the interplay of all these different contributions. Further, when comparing our theoretical onset absorption with the experimental spectrum\cite{kobayashi1983optical}, it is worth recalling that the latter also exhibits a series of higher energy features corresponding to higher energy excitations in ice. The focus of our work is establishing the mechanisms that could lead to emission of photons in the UV-visible range which typically occur from the lowest excited state\cite{Kasha_1950}.

Fig. \ref{Fig2}a shows the distribution of emission energies in the minimum of the first ES PES (C in Fig. \ref{Fig1}a), computed for 100 configurations. The emission energies vary between 4.3-5.5 eV with a significant shift of $\sim$4-5 eV compared to the onset of absorption. An analysis of the geometry and of the electronic orbitals (Fig. S2 in the SI) in B and C on the ES PES shows that along the B $\rightarrow$ C path, molecular dissociation occurs in all the configurations sampled here, except one, for which we observed instead the formation of a pair of Bjerrum defects, after geometry optimization in the excited state. This result will be discussed in detail later in section \ref{BD_sec}. 

Here we focus on the dissociative path. We find that molecular dissociation leads to the formation of a hydronium ion, a hydroxyl radical and an excess electron that in configuration C is delocalized across a six-membered ring close to the the hydroxyl radical (the single particle orbitals involved in the transition are shown in Fig. S2b in the SI). These results suggest that the presence of an electron on water-like wires can stabilize the transient charge-separated state, consistent with the recent findings of Ref. \cite{Tang_2025}, which attributed the charge-transfer exciton peak ($\sim$8 eV) in the absorption spectrum of ice Ih to collective excitations in hydrogen-bonded water wires. The excess electron, initially delocalized along the six-membered rings in C, recombines with a hydroxyl radical to yield ionic photoproducts H$_3$O$^+$ and OH$^-$, during the emission of a photon from C to D. 

To understand in detail the electronic structure of the species associated with the relaxation process, B $\rightarrow$ C, and then the emission of a photon from configurations C $\rightarrow$ D, we computed the linear-response differential density, $\Delta \rho^{(x)}(\textbf{r})$ (see Fig. \ref{Fig2}b), which corresponds to the difference between the electron density of the system in C and D (see the \textit{Materials and Methods} section for details). $\Delta \rho^{(x)}(\textbf{r})$ is reported in Fig. \ref{Fig2}b for one representative configuration from the optimized ES PES.
The electron depletion, shown in blue, is localized around OH, while the electron accumulation, shown in yellow, is found in proximity of H$_3$O and surrounding H$_2$O molecules. During the emission (C $\rightarrow$ D) process, the electron density shifts from the yellow regions to the blue regions. This redistribution indicates that the photon emission is accompanied by the following charge reorganization process: $\text{H}_3\text{O}^+ + \text{OH}^. + \text{e}^- \rightarrow \text{H}_3\text{O}^+ + \text{OH}^-$.

To understand the origin of the spread in emission energies ($\sim1.2$ eV), we examined the distortions of the hydrogen bonding network around the photoproducts \cite{BENOIT199923,Piaggi18102021}. Specifically, we determined the distances (referred to as d1, d2, and d3) between the oxygen atom of the OH$^.$ radical and its three nearest oxygen neighbors as a function of the computed emission energies. We then focused on the shortest of these distances (d1), and distinguished groups of nearest neighbor (NN) oxygen atoms based on whether they belong to H$_3$O$^+$ or an H$_2$O molecule. Our results, presented in Figs. S3 and S4 reveal that both the separation between photoproducts and the OH$^.$-H$_3$O$^+$ distance are key factors in determining the onset of emission energies, which become more red-shifted as the distance decreases.

The photochemistry resulting in the dissociation of an O-H bond of water has been extensively discussed in the literature\cite{Hans_2017,Yuan_2011,Stetina_2019,Dutuit_1985,kivimaki_2006,Acocella_2012}. Specifically, the photodissociation in the first ES of water has been observed in calculations using a repulsive potential energy surface that leads to the transformation of a water molecule into a hydrogen atom or excess proton, leaving behind OH radicals or also an excess electron\cite{Yuan_2011}. More recent, ES simulations of water nanodroplets have reported photoproducts similar to those identified here\cite{Stetina_2019}. Our simulations, however, are the first to capture this dissociation process in condensed-phase ice (investigated with calculations with periodic boundary conditions) using highly accurate electronic structure methods. As we will see shortly, the underlying mechanisms involving the generation and localization of the radicals and excess electrons, sensitively depend on the presence of defects in the hydrogen-bond network.

\subsection{Vacancy Defects}

Many characteristic properties of ice arise from imperfections in the crystal \cite{petrenko1999physics,deKoning2020}; vacancies, for example, are a class of point defects that play an important role in driving diffusion processes in ice and in stabilizing excess electrons \cite{Bhattacharya_2014,deKoning_2016}. Hence, it is interesting to investigate how vacancies may affect the photophysical and photochemical properties of ice. As in the DF case, we computed the absorption spectra for 100 lattice configurations, considering the lowest 100 excitations. We find similar optical features in all configurations (see Fig. S5), and a narrow distribution of absorption onset energies, that are red-shifted by about 0.6 eV compared to the DF case (see Fig. \ref{Fig2}c). 

The presence of a vacancy in the ice crystal gives rise to an occupied defect state in the gap localized on the H$_2$O molecules surrounding the vacancy (see Fig. S6a). The lowest absorption transition occurs from this defect state, located about 0.3 eV above the VBM (see the density of states (DOS) plot in Fig. S7b), to the conduction band. Hence, the onset of absorption in the crystal with a vacancy is red-shifted relative to that of pristine ice.
 
The presence of a vacancy also induces a red-shift in the emission onset energy distribution (green histogram in Fig. \ref{Fig2}c). As in the case of DF, we find a dissociative path in the ES PES that leads to the formation of $\text{H}_3\text{O}^+$, $\text{OH}^.$, and $\text{e}^-$ photoproducts, as shown by the linear-response differential density (Fig. \ref{Fig2}d) and orbital analysis (Fig. S6b in the SI). However, at variance from the DF crystal, we find that in the ES (C) the electron is not delocalized on water wires, but rather localized within the vacancy (hence the emission energy red-shift, relative to DF ice) creating a hydrated electron \cite{Weissmann_1973,Herbert_2017}. Similar to the DF case, upon emission of a photon (from C to D), the electron localizes on an OH group, which is located in the vicinity of the vacancy.

The localization of the excess electron within the cavity is shown in Fig. \ref{Fig2}d; the electron is stabilized by the surrounding water molecules, whose protons reorient towards the electronic localized charge. The electron depletion region (blue area in Fig. \ref{Fig2}d) is well localized on OH$^.$. While it is beyond the scope of the current study to investigate the kinetics of the electron-hole recombination, the presence of the vacancy, which serves as a trapping site for the electron, is expected to reduce the probability of non-radiative decay \cite{AiMinhua2020,Cai2023}.

\begin{figure}[H]
\centering
\includegraphics[width=\linewidth]{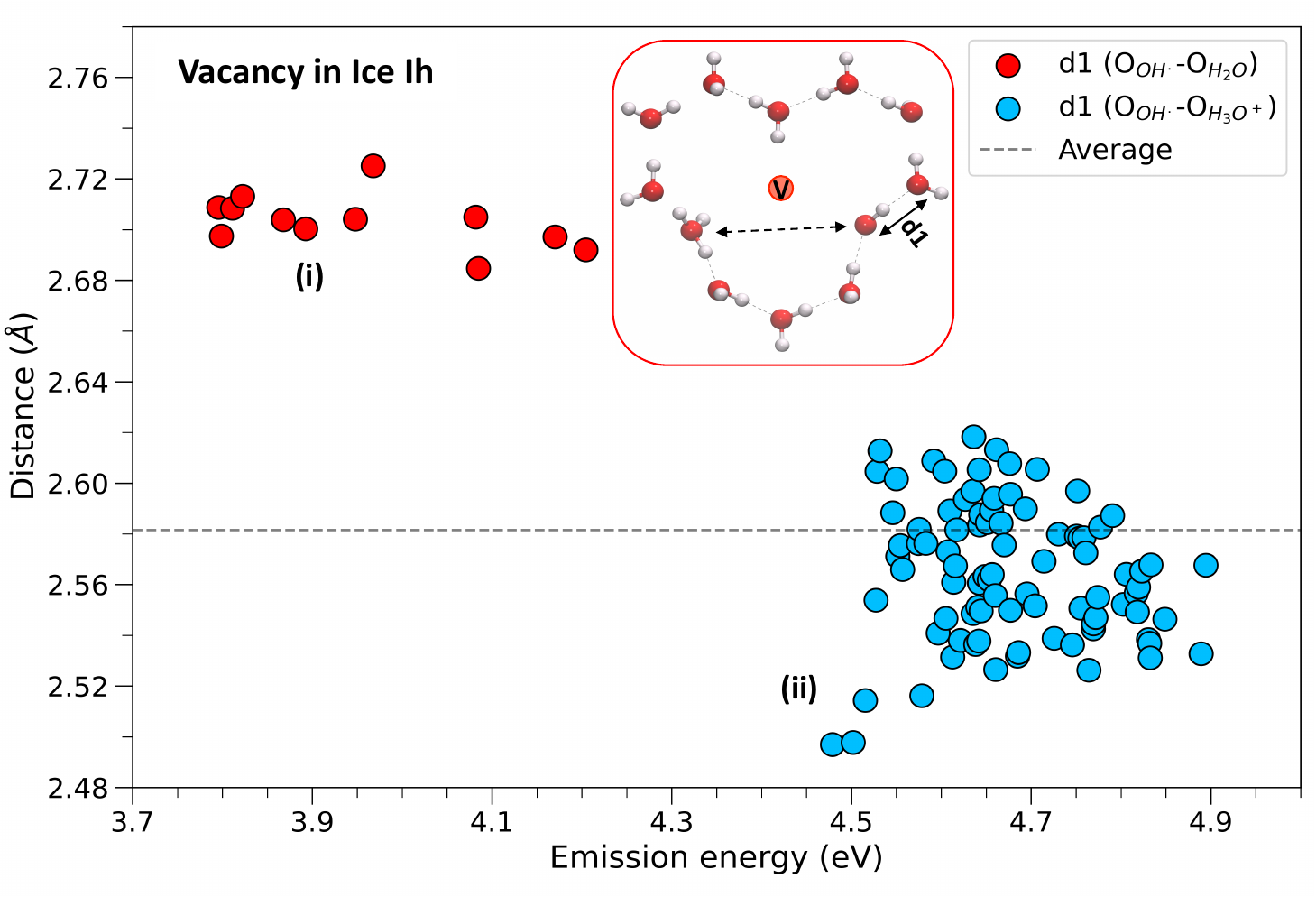}
\caption{Distances between the oxygen belonging to an hydroxyl OH$^.$ and its nearest oxygen neighbor (NN) are shown as a function of the emission energy of disordered ice with a vacancy defect, for 100 configurations. NN oxygen atoms belonging to H$_2$O (group (i)) or H$_3$O$^+$ (group (ii)) are represented with red and cyan dots, respectively. An example of the region near the vacancy for a configuration in group (i) is shown in the inset.}
\label{Fig3}
\end{figure}

As in the DF case, we analyzed the local hydrogen-bonding environment around the hydroxyl radical in the presence of the vacancy by computing the distances (d1, d2 and d3) between the O atom of OH$^.$ and its three nearest oxygen neighbors, as shown in Fig. S8. This structural characterization allows us to probe whether the local coordination of the OH radical correlates with the emission energy. In Fig. \ref{Fig3}, we show the shortest O$_{OH^.}$-O distances (d1) as a function of the emission energy. We observe two distinct groups of NN: those where the O atom belongs to a H$_2$O molecule (red dots, group (i)) and those where the oxygen belongs to H$_3$O$^+$ (blue dots, group (ii)). The former group, with longer d1 distances, has smaller emission energies (smaller than 4.3 eV, starting at 3.8 eV); the latter group, with shorter d1, has larger emission energies, centered at 4.7 eV. Interestingly, for group (i), our ES geometry optimizations reveal that the hydronium ion diffuses away from the hydroxyl radical, with both species remaining near the cavity, as shown in the inset of Fig. \ref{Fig3}.

We note that a larger separation between H$_3$O$^+$ and OH$^-$ correlates with a destabilization of the GS (up to $\sim$ 1.0 eV relative to the average configuration), due to the weakened electrostatic interactions between the two ions, and with a modest stabilization of the ES ($\sim$0.15-0.2 eV). In contrast to group (i), in group (ii) the OH$^.$ and H$_3$O$^+$ are always NNs. Here, a slight red-shift is observed when the two products are in closer contact (shorter d1), although this effect is less pronounced compared to that induced by the separation in group (i). Notably, the photoproducts are formed near the vacancy for configurations (ii) as well.  

\subsection{Ionic Defects}

In addition to vacancies, experimental ice samples contain ionic products of water autoionization, namely the proton and hydroxide ions, which are key players in controlling its conductivity \cite{deKoning2020}. In addition, based on the experiments of Refs. \cite{quickenden1985uv,matich1993oxygen}, where ice samples were photo-irradiated for several hours, we expect a concentration of photoproducts to accumulate within the lattice and hence play a role in determining the photophysical properties of ice. To study one of the major photoproducts, we generated an ice model where we introduced in the GS structure a negatively charged hydroxide ion (OH$^-$) and explored its influence on the optical properties of the crystal.   

A single OH$^-$ was introduced in 40 DF configurations, and in 40 configurations containing a vacancy. For each configuration, we first computed the absorption energies of the 100 low-lying excitations. The results (Fig. S9 in the SI) exhibit consistent features across the ensemble of configurations. 

\begin{figure}[H]
\centering
\includegraphics[width=\linewidth]{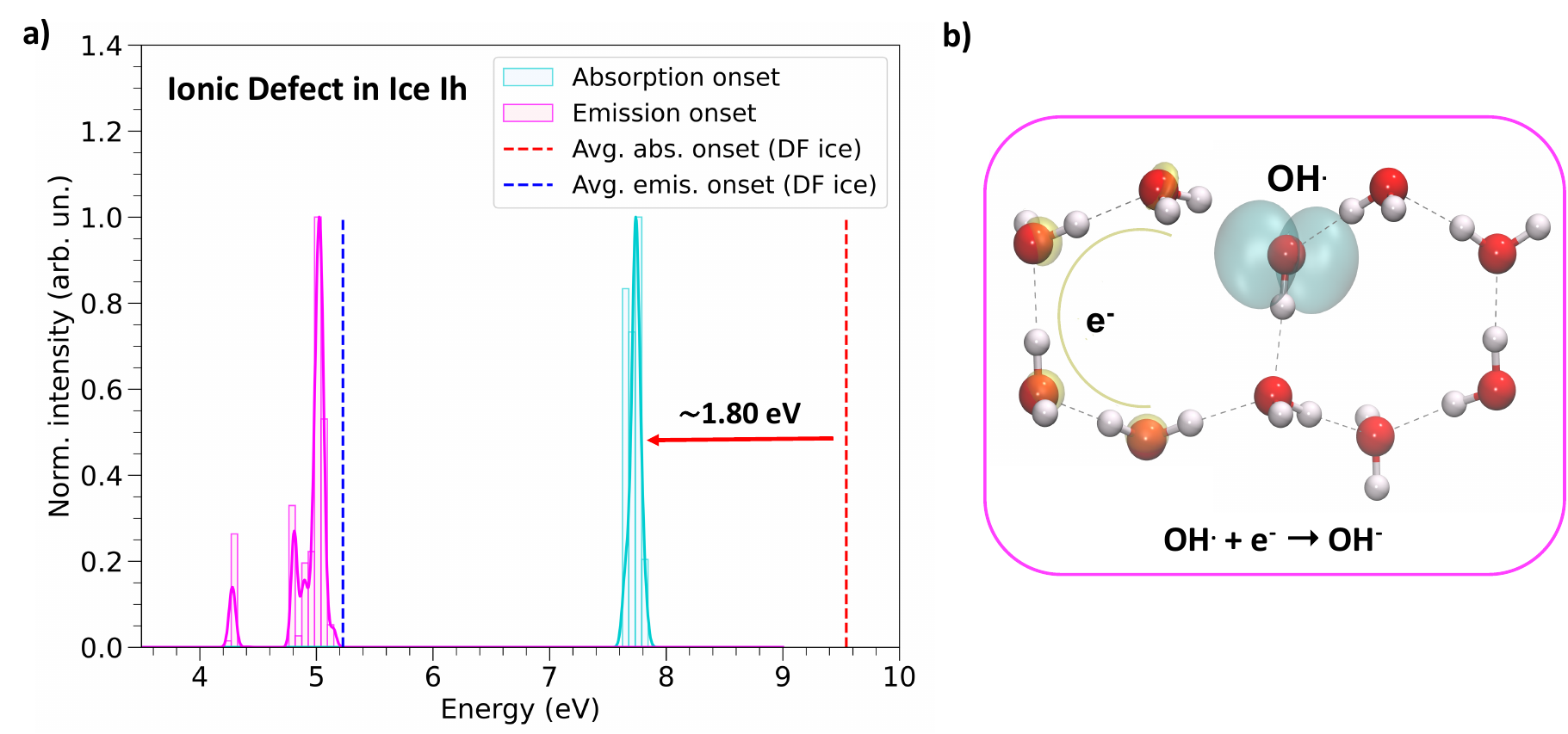}
\caption{Distributions of absorption (cyan) and emission (magenta) onset energies for 40 ionic defect (OH$^-$) configurations of disordered ice Ih are shown in panel a. Solid lines represent Gaussian kernel density estimation (KDE) curves fitted to the corresponding normalized histograms (see \textit{Methods}). Average (Avg.) values of the absorption (abs.) and emission (emis.) onset of the defect-free (DF) ice model are shown as red and blue dashed vertical lines, respectively. An example of unrelaxed differential density is displayed in panel b, showing electron depletion and electron accumulation in blue and yellow, respectively.}
\label{Fig4}
\end{figure}

Fig. \ref{Fig4}a clearly shows the substantial impact of ionic defects on the distribution of the onset absorption energies (cyan histogram). Compared to the DF model (red dashed line), the distribution is significantly red-shifted, with a mean value around 7.7 eV--approximately 1.8 eV lower than that of the DF case (9.5 eV). The presence of OH$^-$ in the lattice gives rise to a defect level about 2.1 eV above the VBM of the crystal (see the DOS plot in Fig. S7c) and the lowest absorption occurs from this level, localized on the defect and surrounding water molecules, to a delocalized state hybridized with the conduction band minimum (see Fig. S10a).

The introduction of OH$^-$ leads to a moderate red-shift of the onset of emission energies compared to the case of DF ice (blue dashed line). The emission mechanism (C $\rightarrow$ D in Fig. \ref{Fig1}a) is governed by the formation of OH$^.$ and an e$^-$ in the ES (configuration C), followed by the localization of the electron on OH upon emission of a photon (C to D). The formation of H$_3$O$^+$ is not observed in this case. The linear-response differential density (Fig. \ref{Fig4}b) shows a region of electron depletion (blue) localized on OH and regions of electron accumulation on nearby H$_2$O molecules. The nature of the single particle orbitals involved in the transition (Fig. S10b in the SI) is also consistent with these regions of charge accumulation and depletion. 

When both a vacancy and an ionic OH$^-$ defect are introduced in the unit cell, we find that during optimization of the ground state geometry, the anionic species--though placed randomly in the lattice in its initial configuration--migrates toward the surface of the cavity to minimize distortions in the bulk hydrogen-bond network. The resulting absorption and emission onset energy distributions (see Fig. S11a) closely resemble those in Fig. \ref{Fig4}, with the main difference being that, in the ES, the excess electron (e$^-$) localizes within the cavity, in close proximity to the OH$^.$ radical, as revealed by the linear-response differential density (see Fig. S11b).
  
\subsection{Bjerrum Defects}
\label{BD_sec}

As mentioned earlier, one of the ES optimizations carried out for DF ice led to the formation of a configuration closely resembling topological coordination defects commonly referred to as Bjerrum defects (BDs) \cite{Bjerrum1952,SCHULSON2001} (see Fig. \ref{Fig1}e) which violate the ice rules. This results in two distinct types of defects along the O-O bond. One shares similar features with a so-called L-defect, which is characterized by the absence of a proton along the hydrogen-bond, and a second one closely resembles a D-defect, which involves the presence of two protons along the O-O bond. We find that in the presence of Bjerrum defects, no chemical reactions occur during the emission process and the emission energy is below 4.5 eV.

\begin{figure}
\centering
\includegraphics[width=0.8\linewidth]{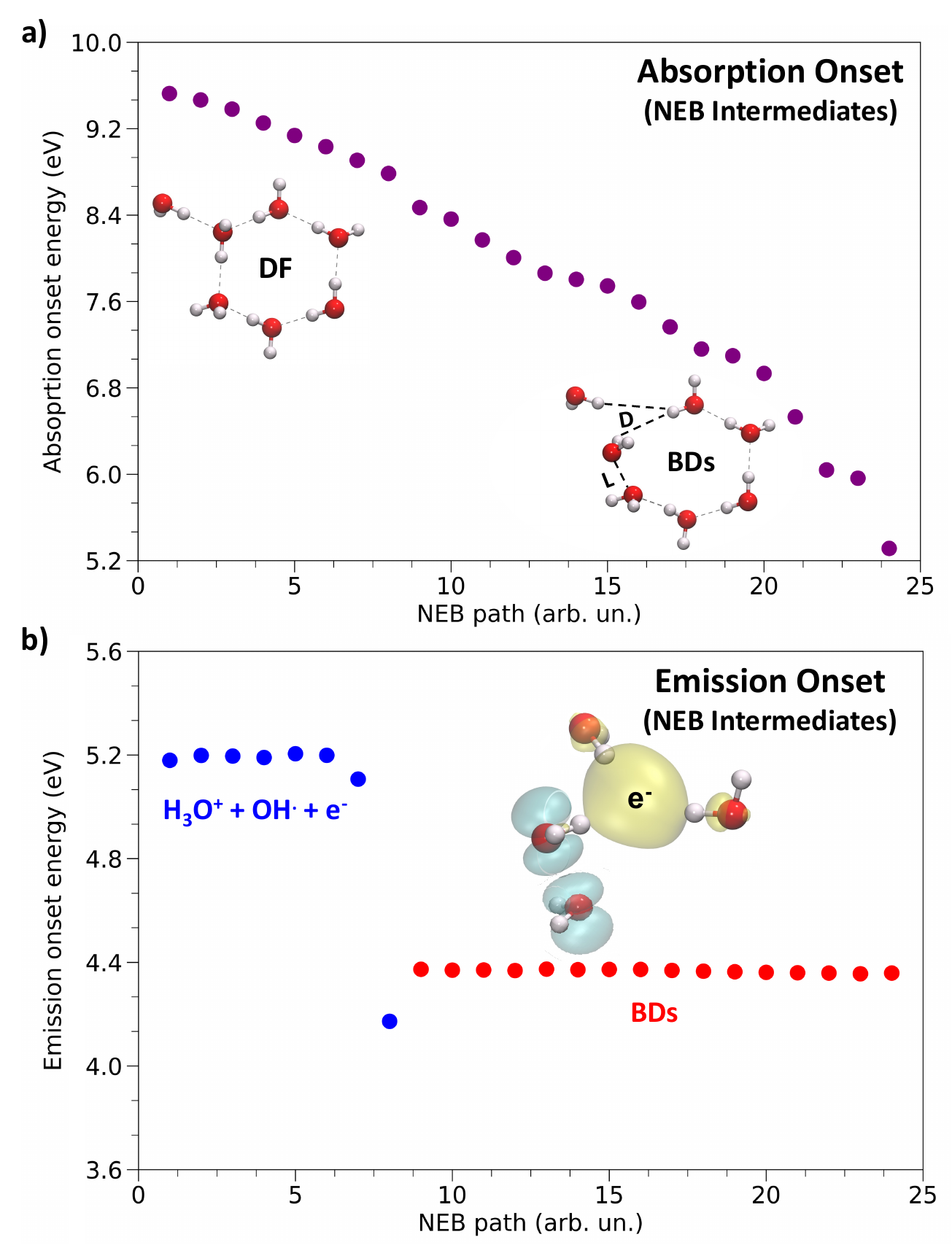}
\caption{Absorption onset energy values (a) calculated for intermediate configurations determined by nudged elastic band (NEB) calculations, from defect-free(DF)-like structures to Bjerrum defects (BDs) structures. An example of DF and BDs moieties are shown in the same panel. The corresponding emission onset energy values are shown in panel b. Blue and red dots represent configurations optimized in the excited state, where OH$^.$, H$_3$O$^+$, and e$^-$ (DF-like) and Bjerrum defects are found, respectively. An example of unrelaxed differential density for one of the BDs-like conformations is shown in the same panel. Electron depletion and electron accumulation are shown in blue and yellow, respectively.}
\label{Fig5}
\end{figure}

To examine in detail the role of BDs in the optical processes of ice, we examined a possible pathway leading to the their creation in the ice crystal. We performed nudged elastic band (NEB) \cite{NEB_2008} calculations using the initial DF configuration and the corresponding BDs structure formed in the ES as the initial and final states, respectively (see the \textit{Materials and Methods} section). Our main motivation for constructing this NEB path on the GS PES is rooted in the fact that the Bjerrum-like configuration we observe represents only one possible geometry. We thus wanted to explore how intermediate configurations along a putative path from water molecules obeying ice-rules, to the Bjerrum-like defect, affect optical properties such as excitation and emission energies. The NEB energy profile (see Fig. S12 in the SI) shows a monotonic increase in energy along the path, which appears to imply that forming the BD configuration incurs an energy of approximately 3.5 eV, relative to the GS. Note that this large energetic penalty originates from the fact that the ES geometry involves not only the L and D pair, but also specific reorganization of the surrounding water molecules that stabilize an excess electron in a nearby cavity, thus leading to the formation of an extended D-like defect (see Fig. \ref{Fig5} and Fig. S13). It is also worth mentioning that L and D defects in real samples of ice are likely generated at interfaces with much smaller formation energies than those found here, before migrating into the bulk\cite{Watkins2010}.

Fig. \ref{Fig5}a shows the decrease in absorption energies during the formation path of the BDs. Once Bjerrum defects are formed, localized defect states appear within the band gap of ice, as shown in the DOS plot in Fig. S7d. The lowest-energy absorption transition occurs between the highest occupied defect state, positioned $\sim3.3$ eV above the VBM, and an unoccupied state localized on the BDs and nearby H$_2$O molecules (see Fig. S14a). The red-shift of the absorption onset relative to DF ice is about 3 eV, not dissimilar to the one found experimentally between short and long absorption exposure; this result indicates that after long UV exposure, BDs may indeed be present in the crystal and give rise to transitions with much lower energy, relative to those of vacancies and ionic defects. We expect however, that finite temperature and NQEs will likely lead to a distribution of BD topologies in the hydrogen-bond network of ice and thus these defects may play a role in the optical properties associated with both short and long UV exposure.

From the NEB reaction path shown in Fig. \ref{Fig5}a, we conducted geometry optimizations in the ES. Our results, shown in Fig. \ref{Fig5}b, reveal two groups of configurations: OH$^.$, H$_3$O$^+$, and e$^-$ (shown in blue), as found in DF ice, and Bjerrum defects (shown in red). The optimized ES configurations for the former (blue dots) are consistent with the previous energetics observed in the DF models--the configuration presenting a red-shifted emission energy of 4.2 eV corresponds to significantly distorted lattice configuration. As shown in Fig. S15a, in this distorted configuration, the OH radical moves away from the hexagonal ring, disrupting the hydrogen bond network, and the OH$^.$ distances and angles with the first three oxygen neighbors (Fig. S15b in the SI) reveal a significant deviation from tetrahedrality.

The linear-response differential density for a representative configuration of the BD group, displayed in red in Fig. \ref{Fig5}b, reveals distinct regions of electron accumulation and depletion along the O-O bonds associated with the D- and L-type Bjerrum defects, respectively. Therefore, in the presence of BDs, the photoluminescence arises from localized excitations on water molecules belonging to the Bjerrum defects, as confirmed by the orbitals involved in the transition (Fig. S14b), exhibiting electron densities strongly localized around the L- and D-type Bjerrum defects.

Having examined all the energetic changes for the different ice systems with and without defects, it is interesting to comment on the oscillator strengths associated with the absorption and emission onsets across the different structural models (see Fig. S16 in the SI). In the case of absorption, defect-free and vacancy-containing models exhibit broader distributions of oscillator strengths, extending to relatively high values, relative to those found for other defects. In contrast, ionic and Bjerrum defect models show narrower distributions shifted toward oscillator strengths with lower values, corresponding to $\sim83\%$ and $\sim33\%$ of the defect-free maximum peak. In the case of emission, the defect-free model shows the highest oscillator strength values. On the other hand, the models containing defects from the start yield weaker transitions. It should be stressed here that due to the ensuing photochemistry on the excited state, the identity of the original GS structure is not the only factor determining the transition strength. The vacancy-containing system shows a broader emission oscillator strength distribution than the other defects: while the majority of values falls below a value that corresponds to $10\%$ of the DF maximum, a few configurations exhibit peaks in the range of $\sim40\%$ of the maximum, and a small peak reaches values comparable to those found for the pristine case. Ionic defects remain below values corresponding to $\sim30\%$ of the DF maximum, while Bjerrum defects exhibit the most suppressed emission, with all values below $\sim4\%$ of the pristine case. Thus, based on our findings it appears that the emission from the hydronium, hydroxyl radical and excess electron photoproducts, yields the largest oscillator strength.

\section*{Discussion}

\begin{figure}
\centering
\includegraphics[width=\linewidth]{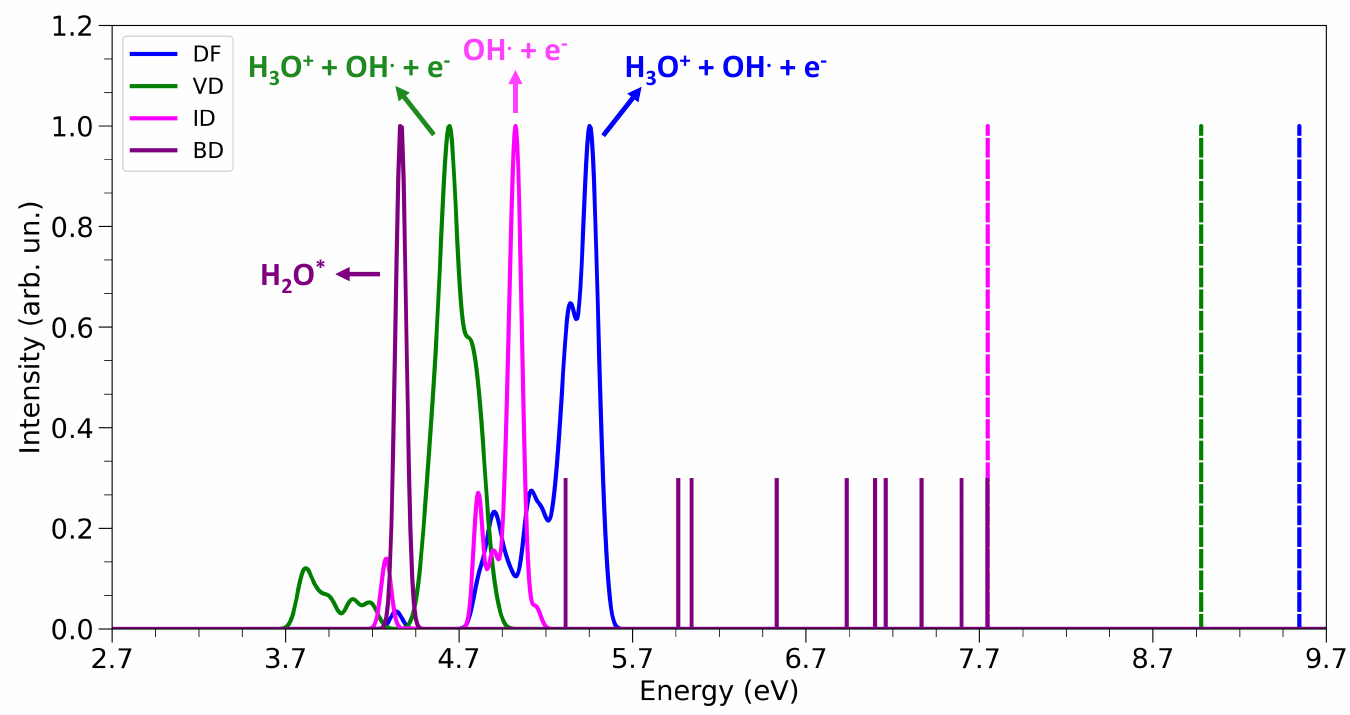}
\caption{Computed absorption end emission onset energies for defect-free (DF, blue), vacancy (VD, green), ionic (ID, magenta), and Bjerrum (BD, purple) defect models. Vertical dashed lines indicate the averaged absorption onset energies for the DF, VD, and ID models. Vertical solid lines correspond to the absorption onset energies of the final ten BD-like configurations obtained from nudged elastic band calculations (see Fig. \ref{Fig5}). Gaussian KDE curves, fitted to the corresponding normalized histograms of the emission onset energy distributions, are shown as colored solid lines for each model. Photoproducts associated with each model are annotated in the figure.}
\label{Fig6}
\end{figure}

In this theoretical contribution, we have taken an important first step to elucidate, with cutting-edge electronic structure methods, the photophysics and photochemistry of ice, highlighting the important role of different defects. Comparing absorption and emission processes in pristine, defect-free ice to those in defective crystals explicitly incorporating vacancies, ionic and Bjerrum coordination topologies, we demonstrate the significant influence of all these defects, providing new perspectives on how to interpret the available absorption and emission experiments, as well as cultivating the way for new measurements.

To the best of our knowledge, high resolution experiments that probe the UV absorption of ice and the resulting photoproducts are scarce. Kobayashi showed that for short exposures to UV light, the first absorption band of hexagonal ice occurs at $\sim$8.7 eV, as seen in the cyan curve in Fig. S17 \cite{kobayashi1983optical}, with an onset at 7.7 eV. Quickenden and co-workers reported the time-evolution of the excitation spectra upon continuous UV-irradiation at 4.8 eV after several hours \cite{quickenden1985uv,matich1993oxygen} - they observe a broad peak centered at 5.6 eV, and an onset more than 1 eV lower (yellow to red curves in Fig. S17). Furthermore, upon excitation at 5.64 eV, in Ref. \cite{quickenden1985uv} they observed luminescence spectra with a peak at 3.65 eV (black curve in Fig. S17). It is also worth noting that in the Quickenden experiments\cite{matich1993oxygen,langford2000}, by exciting at 4.8 eV, again under continuous UV irradiation, the emergence of a long-lived phosphorescence band at 2.95 eV is observed (see Fig. S17). Our current calculations only account for the excitations involving the lowest lying singlet states, hence we cannot make any definitive claims about the origin of this emission peak, which would require the calculation of triplet states. Nonetheless, it is plausible that excitation at 4.8 eV initially populates a singlet state that subsequently undergoes inter-system crossing to a triplet state. Other experiments conducted in the field of planetary science, have shown that UV-irradiation can induce phase transitions in ice possibly catalyzed by the creation of defects \cite{Kouchi_2021}. 

The theoretical findings emerging from our work on onset absorption and emission energies for different ice models are summarized in Fig. \ref{Fig6}. Our results indicate that defects, including vacancies (vertical dashed green line), hydroxide ions (vertical dashed magenta line) and Bjerrum defects that violate the ice-rules (vertical solid purple lines), all contribute to the low energy absorption tail of the spectrum for short UV exposure. On the other hand, the origin of the excitation spectra under long UV exposure is complicated by the fact that the optical spectra may originate from the interplay of features due to intrinsic defects and features generated by photoproducts. Assuming that NQEs can downshift onset absorption energies by approximately 1 eV \cite{Berrens_2024}, Bjerrum-like defects may be the primary species associated with the growth of the peak at 5.6 eV under long exposure. Moreover, it is reasonable to speculate that intrinsic defects may also contribute to the 4.8 eV excitations to singlet states, which can subsequently undergo internal conversion to the triplet manifold.

Our computed emission energies fall within the range of measured ones and arise from a variety of products, including H$_3$O$^+$, OH$^.$, e$^-$, and excited H$_2$O. In all the different model systems investigated here, the emission energies are significantly red-shifted by $\sim$3-5 eV compared to the absorption onset. Interestingly, the model including a vacancy leads to the lowest energy peaks in the range 3.7-4.3 eV (green curve in Fig. \ref{Fig6}). Similarly, a more intense band in the emission onset distribution for the ionic defect model (magenta curve) and the entire distribution for the Bjerrum defects (purple curve) appear within the same energy range. Again, it remains an open question to evaluate how much NQEs will shift the onset emission energies. 

It is worth noting that while our calculations for the DF model indicate that excitation above $\sim$8.5 eV should lead to emission between 4.2 and 5.6 eV, to our knowledge, experiments using comparable excitation energies in ice Ih have not yet been reported. It is very likely that using far UV excitation energies opens up non-radiative decay channels, as observed in liquid water\cite{Yuan_2011,Stetina_2019}. Interestingly, an experimental work by Winter and Knie showed that soft X-Ray irradiated liquid water displays optical fluorescence, including a broad emission band between 3.6 to 7.3 eV\cite{Hans_2017}, suggesting that higher-energy excitations can still give rise to radiative processes under certain conditions.

As alluded to earlier, the peak at 5.6 eV observed in the long-exposure UV may arise from the presence of photoproducts in the ice hydrogen-bonded network. The higher-values of onset emission energies for the DF ice which involve the recombination of the electron with the hydroxyl radical leading to the creation of the hydronium and hydroxide ions in the ground-state, are close to the 5.6 eV peak. Therefore, we speculate that the accumulation of these ion pair photoproducts may play a role in tuning the optical excitations in this energy range.

\section*{Future Perspectives}
In this contribution, we have emphasized the role of different types of defects in tuning the optical absorption and emission properties of ice. In the future, it would be beneficial to carry out more systematic experimental studies, starting by repeating the original measurements by Quickenden \textit{et al}., with newer instrumentation and possibly a better control of the temperature. Further, the IR and Raman spectra of ice exhibit several peaks and shoulders whose origins have remained rather poorly understood\cite{Shi_2013}, and additional combined experimental and theoretical studies of all vibrational signatures would be of great interest to characterize defects. We note that Hamm and co-workers have previously reported inhomogeneous broadening in 2D-IR measurements of ice\cite{Perakis_2011} which has been rationalized in terms of proton disorder. It would be interesting to investigate whether similar measurements can be used to probe the presence of defects. An obvious challenge to address will be the signal-to-noise ratio, possibly due to the relatively low concentration of defects. In this regard, terahertz spectroscopy--which has been used to probe lower-frequency vibrational modes in ice for instance\cite{Tao_2024}--could be a more sensitive probe of the presence of defects. In terms of probing specifically ES properties of ice, time-dependent pump-probe spectroscopy may offer specific clues into mechanisms associated with non-radiative decay. Bakker and co-workers for example, used a UV pulse to trigger the ejection of a proton from a photoacid in ice, followed by a transient IR probe to detect the proton transfer mechanism\cite{Timmer_2010}. One could envisage using ultrafast XUV pulses to probe the time-dependent behavior of the electronic ES as has been done recently to investigate conical intersections in various molecular systems\cite{chang_2020,Keefer_2023}.

In addition to detecting defects, assessing their relative populations will likely be important in determining the fate of photoexcited ice. de Koning et al.\cite{deKoning_2007} have built thermodynamic-based models using DFT calculations to rationalize the dielectric properties of ice\cite{jaccard_1959,petrenko1999physics}; in these models the concentrations of different types of defects are optimized as parameters. They find a wide range of concentrations of Bjerrum-like defects ranging between 1 p.p.b to 1 p.p.m , depending on the assumptions of the model, and the temperature\cite{deKoning_2007}. These concentrations fall within the range of nano-to-micromolar. In addition, positron annihilation spectroscopy measurements\cite{Eldrup_1976,Eldrup_1978,Mogensen_1978} have shown the existence of vacancies in ice with similar concentration. Since the absorption of UV photons, and subsequent photochemistry, will likely entail local heating effects, the concentration of defects will likely increase. It would be interesting to probe these defects under photoexcitation using time-dependent spectroscopies such as IR or Raman\cite{Shi_2013,Perakis_2011}.

While the results of our calculations serve as an important step in understanding the photophysics and photochemistry of ice, there remain important theoretical challenges that need to be tackled in the future. Our ES optimizations neglect both finite temperature dynamics and NQEs. Further, photoexcitation will result in non-equilibrium local heating which may also play an important role in activating and altering the dynamical pathways. While we recognize the importance of these effects, we could not yet fully address their complexity and they will be examined in future works. NQEs, through both zero-point energy fluctuations and tunneling, are well known to affect the structural and electronic properties of water and ices \cite{Engel_2015,Chen_2016, Bischoff2021,Berrens_2024}, and, more generally, of solid and liquid systems containing light elements \cite{Sappati_2016,Kundu_2024,Tsuru_2025,Xu_2025}. Here the role of NQEs on the onset of absorption and emission processes have essentially only been inferred from previous estimates of NQEs effects on the absorption energies of liquid water and pristine ice. Another important aspect that will need to be tackled in future studies is that of non-radiative decay mechanisms requiring techniques that combine electron and nuclear dynamics \cite{Barbatti_2011}. 

\section{Materials and Methods}
\subsection*{Ground state geometry optimizations}
We considered 100 proton-disordered defect-free (DF) configurations of ice Ih represented by $(2\times2\times2)$ supercells (128 molecules), generated by the GenIce2 software version 2.2.7.6 \cite{genice_r1,genice_r2}. The density of the samples was set to $0.94$ $g/cm^3$ in order to match that of ice Ih at the temperature of experiments performed at 77-88 K \cite{quickenden1985uv,matich1993oxygen}. Different random seed values were used to ensure the generation of different hydrogen ordering in the various configurations and the total dipole moment of the cell was set to zero. The 100 vacancy defect (VD) configurations and the 40 ionic defect (ID) configurations were obtained by randomly removing one water molecule and one proton, respectively, from the DF supercells. The 40 configurations combining vacancy and ionic defects were obtained by randomly removing an additional proton from the supercells with a vacancy defect. As the ID and combined vacancy–ionic defect models carry a total charge of –1, a uniform compensating jellium background was automatically introduced to ensure electrostatic convergence under periodic boundary conditions, as implemented in the Quantum ESPRESSO package. Finally, the initial Bjerrum defect (BD) configuration was obtained by optimizing a DF structure in the excited state. 

The geometries of the DF, VD, and ID models were optimized in their GS by using density functional theory (DFT) with the Perdew, Burke, and Ernzerhof (PBE) \cite{PBE} exchange-correlation functional, followed by a refinement with the dielectric-dependent hybrid (DDH) functional \cite{DDH_r1,DDH_r2} (exact exchange fraction of 0.58), using the Quantum ESPRESSO package \cite{QE_r1,QE_r2}. The quasi-Newton Broyden-Fletcher-Goldfarb-Shanno (BFGS) algorithm \cite{Broyden1970,Fletcher1970,Goldfarb1970,Shanno1970} was used for the geometry optimizations. The geometry optimizations were carried out until forces were below $3.9\times10^{-4}$ Ry/Bohr. The kinetic energy cutoff for wavefunctions was set to 70 Ry and we used the optimized norm-conserving Vanderbilt (ONCV) pseudopotentials \cite{ONCV}. The Brillouin zone of the supercells was sampled at the $\Gamma$ point.

\subsection*{Optical properties calculations}
Absorption onset energies for the optimized GS structures were obtained using time-dependent hybrid density functional theory calculations within the Tamm-Dancoff approximation (TDA) and the open-source WEST code \cite{WEST_r1,WEST_r2}. We used the same kinetic energy cutoff, pseudopotentials, and exchange-correlation functional (DDH) employed in the GS geometry optimizations. To accelerate the time-dependent hybrid density functional theory calculations, several numerical approximations were adopted, as described in Ref. \cite{YuJin2023}. We used the adaptively compressed exchange (ACE) operator to approximate the exact exchange operator, setting the number of ACE orbitals to four times the number of occupied Kohn-Sham (KS) orbitals. In evaluating the Coulomb integrals, we transformed the occupied KS orbitals into maximally localized Wannier orbitals \cite{gygy_mlwf} and applied an overlap threshold of $10^{-3}$ to truncate pairs of non-overlapping localized orbitals, thereby greatly reducing the number of integrals to evaluate. The molecular orbitals involved in the transition were computed with a post-processing tool available in Quantum ESPRESSO. 

We performed TDDFT optimizations of the first singlet ES as implemented in the WEST code \cite{WEST_r1,WEST_r2}. The geometry optimizations were carried out until forces were below $3.9\times10^{-4}$ Ry/Bohr. The same numerical approximations adopted to accelerate the absorption TDDFT calculations were employed, along with the inexact Krylov subspace approach with a threshold of $10^{-6}$ applied to the residual vector.

The analysis of both absorption and emission onset energy distributions was performed by computing histograms with 100 bins. Each histogram was normalized such that the maximum bin height equals 1. For both absorption and emission, oscillator strengths were used as weights to account for the relative intensity of each transition. To generate smooth representations of the distributions, Gaussian kernel density estimation curves were constructed using the gaussian-kde function from the SciPy library \cite{2020SciPy-NMeth}. The bandwidth parameter was set to either 0.15 or 0.3, depending on the defect model, to accommodate differences in data spread.

We computed the linear-response differential density, $\Delta \rho^{(x)} (\mathbf{r})$, as the change in electron density between the ES and the GS. The ES wavefunction was approximated as a linear combination of single excitations from the $v$-th occupied KS orbital $\varphi_v (\mathbf{r})$ to the $c$-th unoccupied KS orbital $\varphi_c (\mathbf{r})$, with expansion coefficients $A_{vc}$ obtained by solving the TDDFT equations within the TDA. In practice, $\Delta \rho^{(x)} (\mathbf{r})$ was evaluated as
\begin{equation}
    \Delta \rho^{(x)} (\mathbf{r}) = \sum_{v=1}^{N_{\mathrm{occ}}} |a_{v} (\mathbf{r})|^2  - \sum_{v=1}^{N_{\mathrm{occ}}} \sum_{v'=1}^{N_{\mathrm{occ}}} \varphi_v  (\mathbf{r}) \varphi_{v'}^*(\mathbf{r}) \int \mathrm{d} \mathbf{r'} a_{v}^{\ast} (\mathbf{r'}) a_{v'} (\mathbf{r'}),
\end{equation}
where the transition orbital associated with each occupied state is defined as $a_{v} (\mathbf{r}) = \sum_{c} A_{vc} \varphi_c (\mathbf{r})$. The implementation of TDDFT within the WEST code yields the $a_v$ transition orbitals without explicit summations over empty states thanks to the use of the density matrix perturbation theory (see eq. 17 of Ref.\cite{YuJin2023}).

\subsection*{NEB calculations}
A nudged elastic band (NEB) calculation \cite{NEB_2008} was performed using the climbing image scheme as implemented in the Quantum ESPRESSO package \cite{QE_r1,QE_r2}. The DF configuration and the corresponding BDs structure, which was formed in the ES optimization, were used as initial and final state, respectively. The $\text{DF} \rightarrow \text{BDs}$ reaction pathway was discretized into 14 images. Inputs as exchange-correlation functional, energy cutoff, and k-sampling were the same as in the DFT and TDDFT calculations described above. Since the proton rotation that leads to the L-, and D-type defects formation occurs quite rapidly, we refined our approach by performing a NEB calculation by choosing a given image (image 10) as initial state, while the final state was kept unchanged. Both absorption and emission onset energies were then computed for the set of 24 structures obtained by combining the sets of NEB calculations.





\begin{acknowledgement}

MM, GDM, and AH thank the European Commission for funding on the European Research Council Grant HyBOP 101043272. MM, GDM, and AH also acknowledge Consorzio Interuniversitario del Nord-Est per il Calcolo Automatico (CINECA) supercomputing (project NAFAA-HP10B4ZBB2) and MareNostrum5 (project EHPC-EXT-2023E01-029) for computational resources. MG acknowledges the CINECA award under the Italian Super Computing Resource Allocation initiative, for the availability of high performance computing resources and support, and the EuroHPC Joint Undertaking for awarding us access to Leonardo hosted at CINECA, Italy. The development of codes to study excited state properties and the calculations of spectra (YJ, MG, GG) was supported by Midwest Integrated Center for Computational Materials, as part of the Computational Materials Sciences Program funded by the U.S. Department of Energy, Office of Science, Basic Energy Sciences, Materials Sciences, and Engineering Division through Argonne National Laboratory, under Contract No. DE-AC02-06CH11357. This research also  used the resources of the University of Chicago Research Computing Center and of National Energy Research Scientific Computing Center.

\end{acknowledgement}

\section*{Conflict of Interest}
The authors declare no conflict of interest.

\section*{Data Availability Statement}
Simulated data used in this work have been deposited at \url{https://github.com/mmonti00/PNAS_2025_Ice_Defects} (20 October 2025).

\clearpage

\begin{suppinfo}
\beginsupplement

\subsection{Geometrical analysis of the excited state configurations of the defect-free model of ice}
The distances between the oxygen of OH$^.$ (OH radical) and the three nearest oxygen neighbors, labeled as d1, d2, and d3, were computed as a function of the emission energy for the DF configurations, as shown in Fig. \ref{SI_3}. We then considered the relation between the shortest O$_{OH^.}$-O distance (d1) and computed emission energies (Fig. \ref{SI_4}); we considered two cases: one where the nearest neighbor (NN) is a neutral water molecule (red dots), and one where the NN is H$_3$O$^+$ (cyan dots). As shown in Fig. \ref{SI_4}, three groups can be identified: (i) large d1 and low emission energy values, (ii) small d1 and low emission energy values, and (iii) average d1 and average emission energy values. A small transition region connecting groups (ii) and (iii) can also be observed. In group (i), the NN of OH$^.$ is a neutral water molecule, as shown in panel b. In this geometry, associated with the lowest energy in our dataset, H$_3$O$^+$ is not one of the first three neighbors. For the remaining configurations of group (i), H$_3$O$^+$ can be either second or third neighbor. The separation of photoproducts observed in group (i) leads to a red-shift in the emission onset. A similar red-shift can be observed when H$_3$O$^+$ is first neighbor and located in close proximity to OH$^.$ (group (ii)) as shown in panel c. When the OH$^.$-H$_3$O$^+$ distance increases, with the two species still being nearest neighbors, a slightly larger emission energy is obtained (group (iii)). Therefore, our analysis indicates that the distance between photoproducts, as well as a short OH-H$_3$O distance, are responsible for the reduction of the emission onset in the DF ice Ih crystal. 
\newpage

\begin{figure}
\centering
\includegraphics[width=\textwidth]{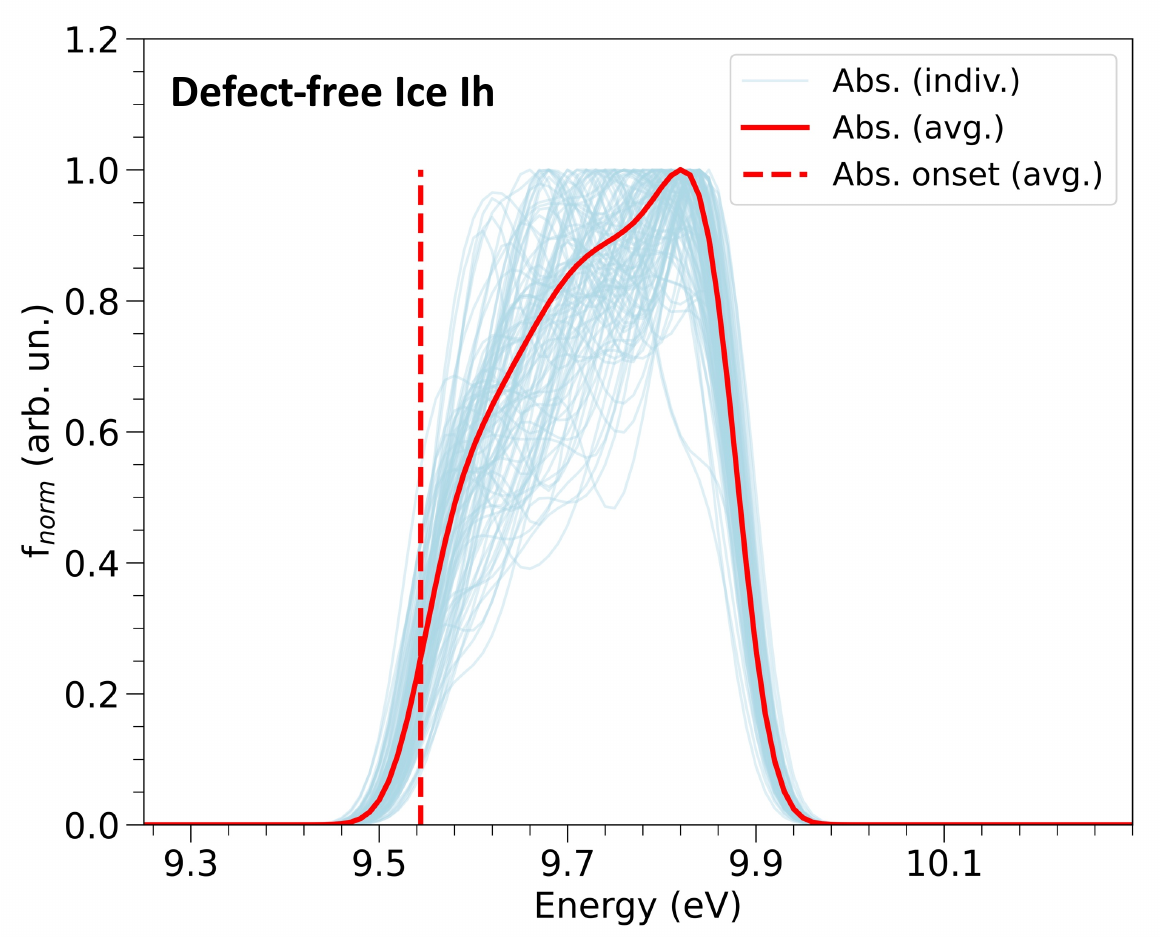}
\caption{Calculated individual (indiv.) optical absorption (Abs.) transitions (solid blue curves) of 100 defect-free Ice Ih configurations. The onset spectrum was obtained by computing the lowest 100 excitations at the time-dependent density functional theory level (see the \textit{Methods} section in the main text for details) and applying a Gaussian broadening with a $\sigma$ of 0.03 eV and a resolution of 0.01 eV. The corresponding average (avg.) absorption spectrum onset is shown as a solid red curve. The absorption onset value, averaged over the 100 configurations, is reported as a vertical dashed red line.} 
\label{SI_1}
\end{figure}

\begin{figure}
\centering
\includegraphics[width=\textwidth]{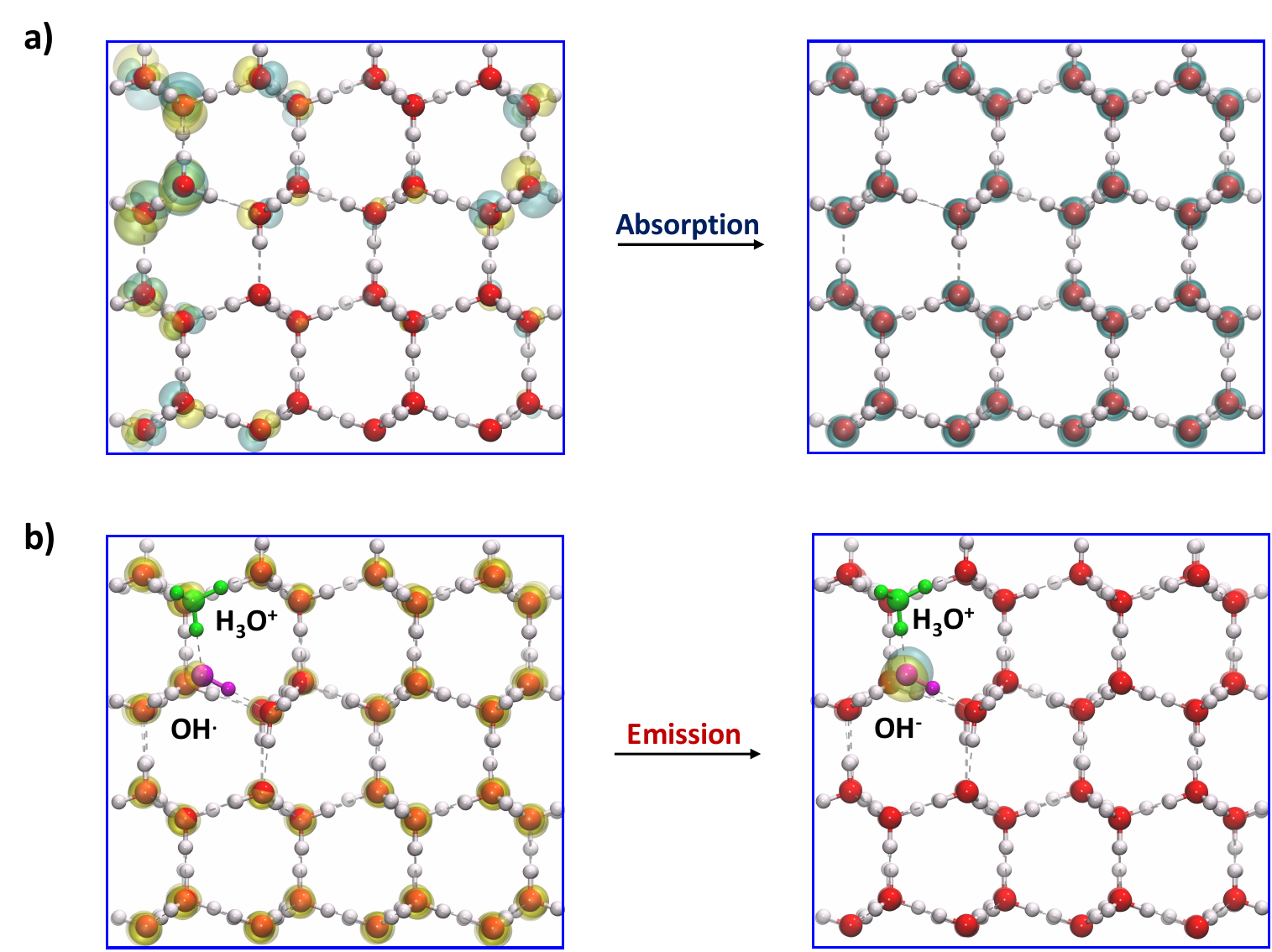}
\caption{The molecular orbitals (MOs) involved in the absorption onset of the defect-free model are shown in panel a. The highest occupied molecular orbital (HOMO, configuration A in Fig. \ref{Fig1} of the main text) and the lowest unoccupied molecular orbital (LUMO, configuration B in Fig. \ref{Fig1}), corresponding to the VBM and CBM of the ice model, are displayed on the left and right, respectively. Orbitals corresponding to configuration C and D of Fig. \ref{Fig1} in the main text, defining the emission onset for the same system are shown in panel b. The hydronium ion (H$_3$O$^+$) and hydroxyl radical (OH$^.$), formed in the excited state, are represented as green and magenta spheres, respectively. In the ground state, the hydroxyl radical recombines with the excess electron to form the hydroxide ion (OH$^-$) which is shown in magenta. Positive and negative lobes of the MOs are shown in yellow and blue, respectively.} 
\label{SI_2}
\end{figure}

\begin{figure}
\centering
\includegraphics[width=\textwidth]{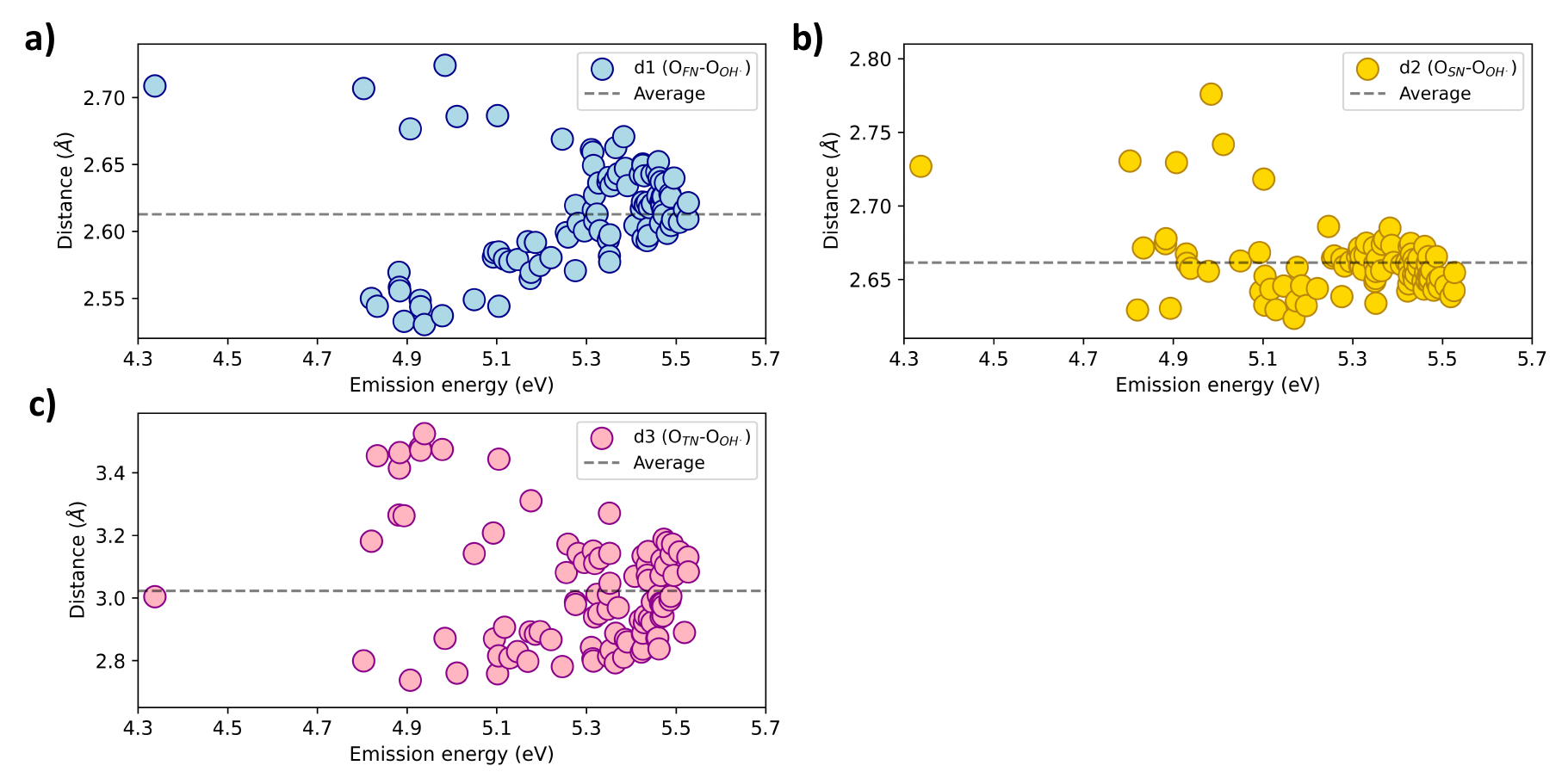}
\caption{Distances between the oxygen atom belonging to OH$^.$ and its first, second, and third oxygen neighbors are shown as a function of the emission energy of disordered defect-free ice Ih, for 100 configurations. Panel a) (light blue dots) shows the distances between the oxygen of OH$^.$ and the first nearest neighbor (d1). Panel b) (gold dots) displays the distances between the oxygen of OH$^.$ and its second nearest neighbor (d2). Panel c) (pink dots) shows the distances between the oxygen of OH$^.$ and its third nearest neighbor (d3). The labels FN, SN and TN in the figure correspond to the first, second and third nearest neighbors.}
\label{SI_3}
\end{figure}
\newpage

\begin{figure}
\centering
\includegraphics[width=\textwidth]{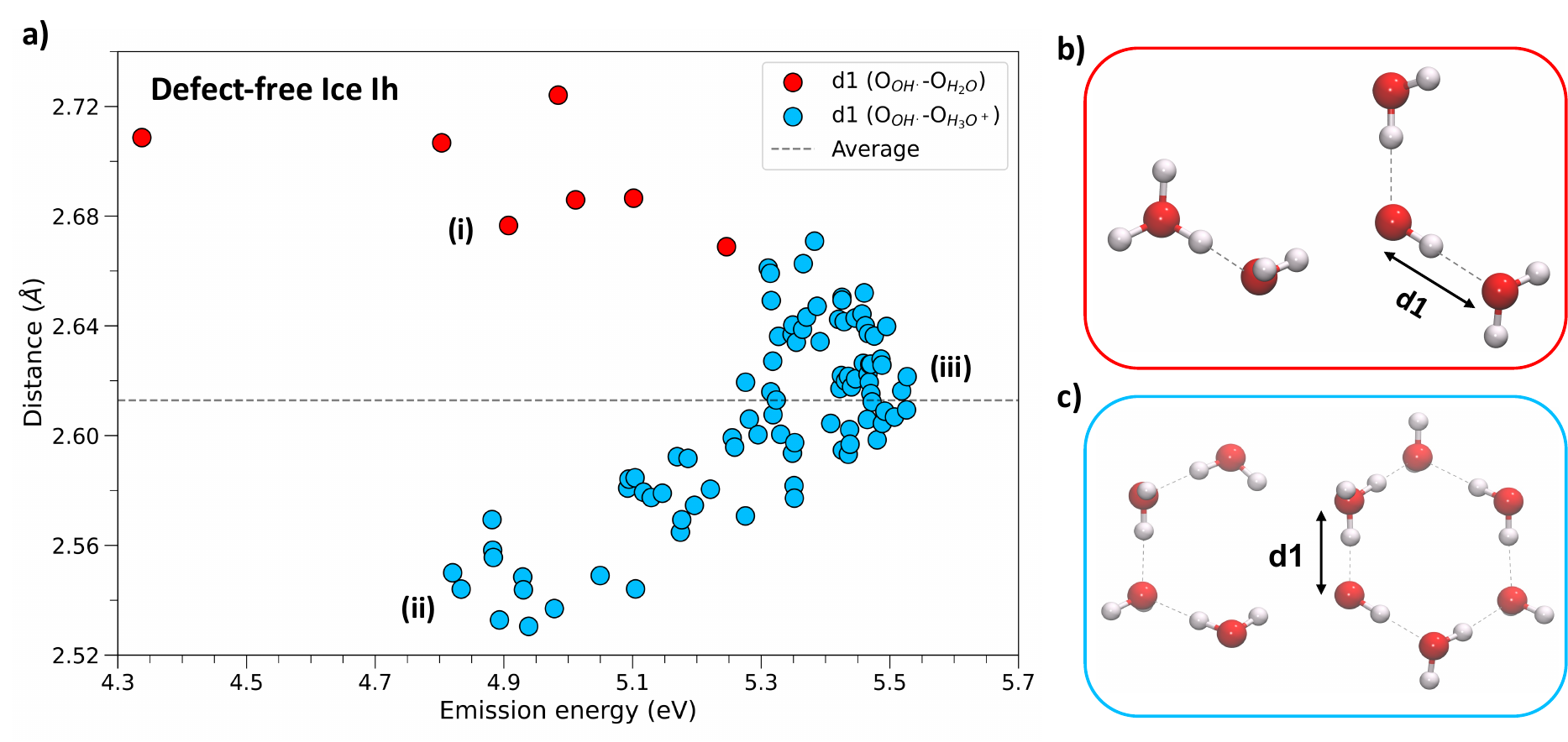}
\caption{Distances between the oxygen atom belonging to an hydroxyl OH$^.$ and its nearest oxygen neighbor (NN) are shown as a function of the emission energy of disordered defect-free ice Ih, for 100 configurations. NN oxygen atoms belonging to H$_2$O or H$_3$O$^+$ are represented with red and cyan dots, respectively. Examples of the region near OH$^.$ for configurations in group (i) and group (ii) are shown in panels b and c, respectively.}
\label{SI_4}
\end{figure}

\begin{figure}
\centering
\includegraphics[width=\linewidth]{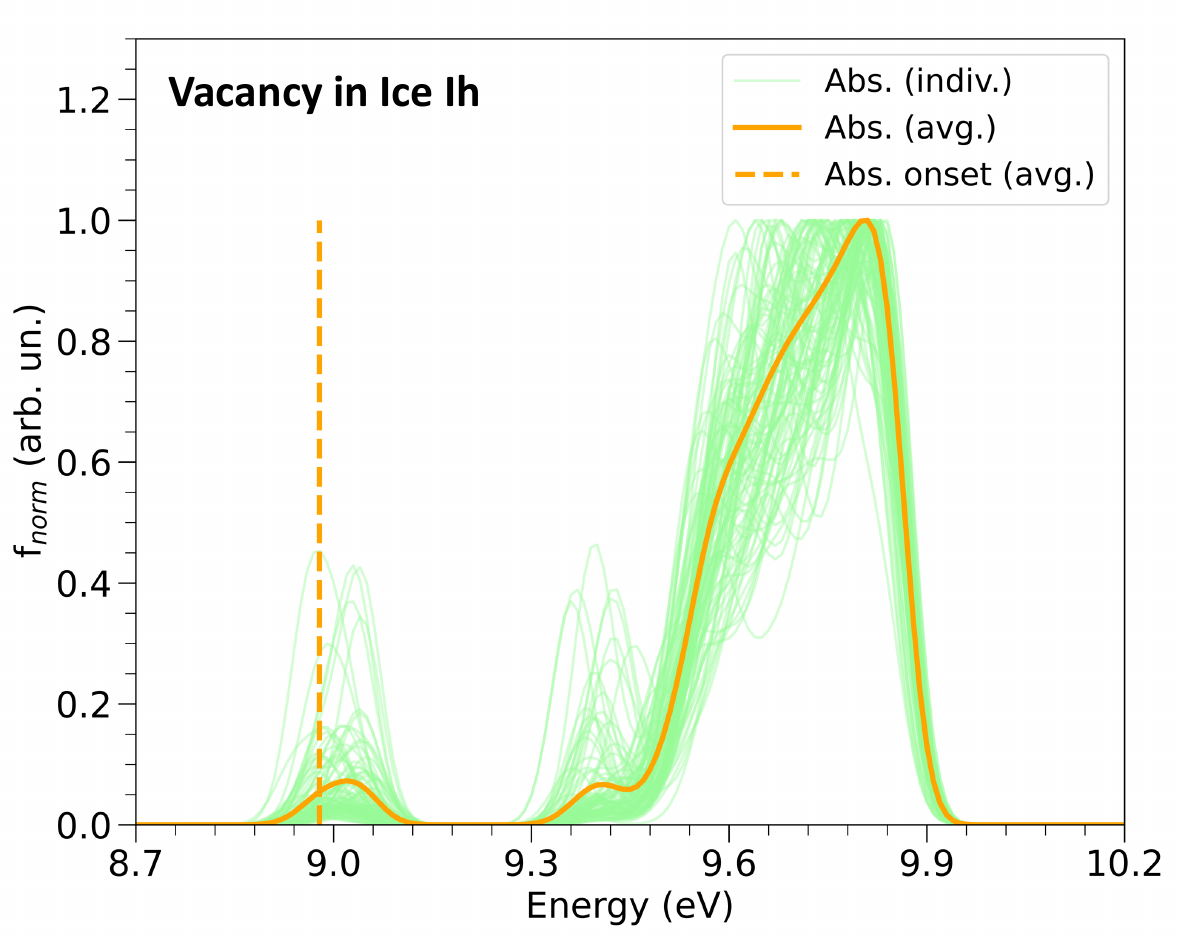}
\caption{Calculated individual (indiv.) optical absorption (Abs.) spectra (solid green curves) of 100 vacancy-containing Ice Ih configurations. The onset spectrum was obtained by computing the lowest 100 excitations at time-dependent density functional theory level (see the \textit{Methods} section in the main text for details) and applying Gaussian broadening with a $\sigma$ of 0.03 eV and a resolution of 0.01 eV. The corresponding average (avg.) absorption spectrum is shown as a solid orange curve. The absorption onset value, averaged over the 100 configurations, is reported as a vertical dashed orange line.} 
\label{SI_5}
\end{figure}

\begin{figure}
\centering
\includegraphics[width=.9\linewidth]{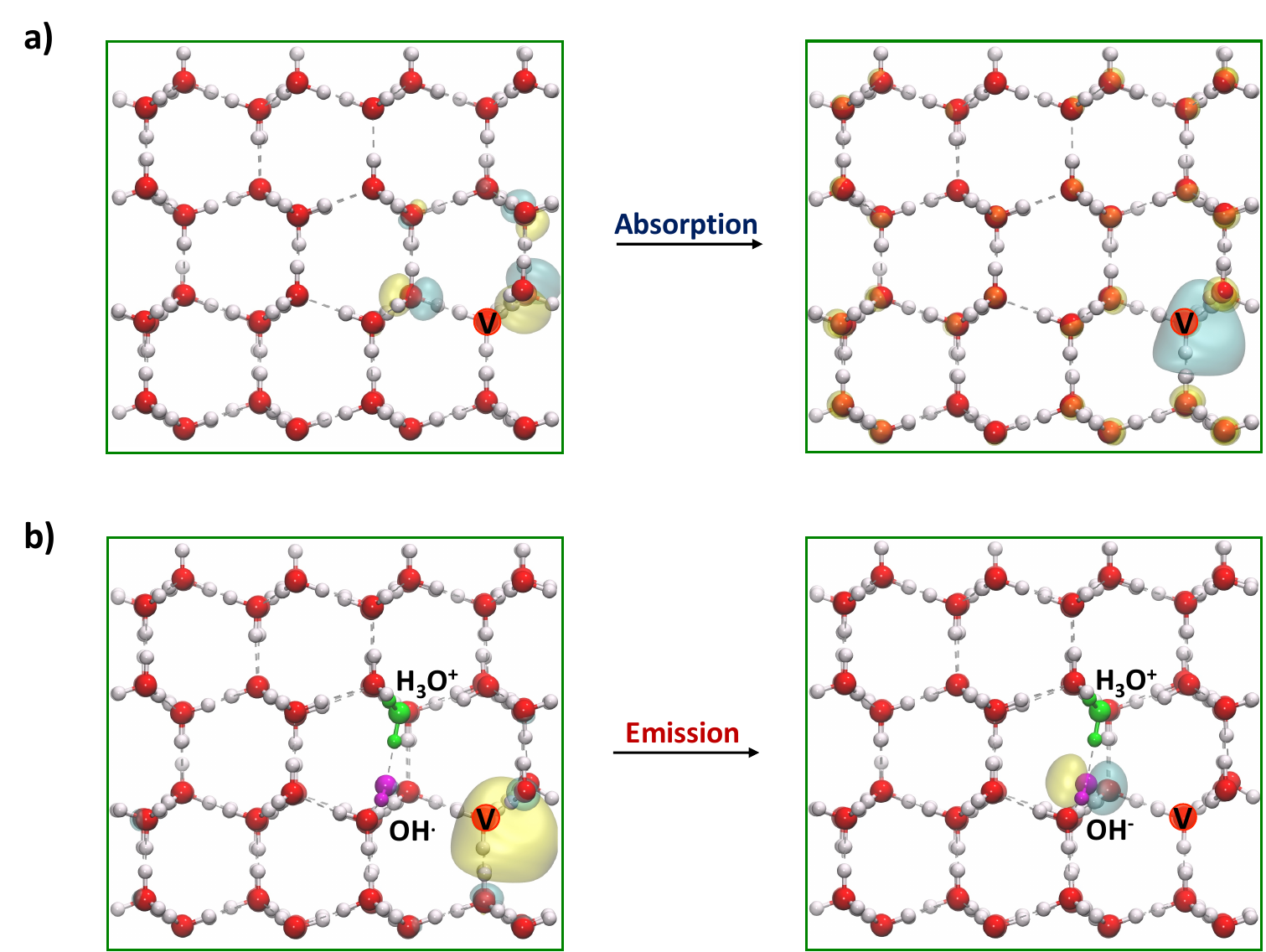}
\caption{The molecular orbitals (MOs) involved in the absorption onset of the ice Ih model with a vacancy (V) are shown in panel a. The highest occupied molecular orbital (HOMO) and the lowest unoccupied molecular orbital (LUMO) are displayed on the left and right, respectively. LUMO (left) and HOMO (right) defining the emission onset for the same system are shown in panel b. The hydronium ion (H$_3$O$^+$) and hydroxyl radical (OH$^.$), formed in the excited state, are represented as green and magenta spheres, respectively. In the ground state, the hydroxyl radical recombines with the excess electron to form the hydroxide ion (OH$^-$) which is shown in magenta. In both panels, positive and negative lobes of the MOs are shown in yellow and blue, respectively. The vacancy site is shown as a red circle.}
\label{SI_6}
\end{figure}

\begin{figure}
\centering
\includegraphics[width=.9\linewidth]{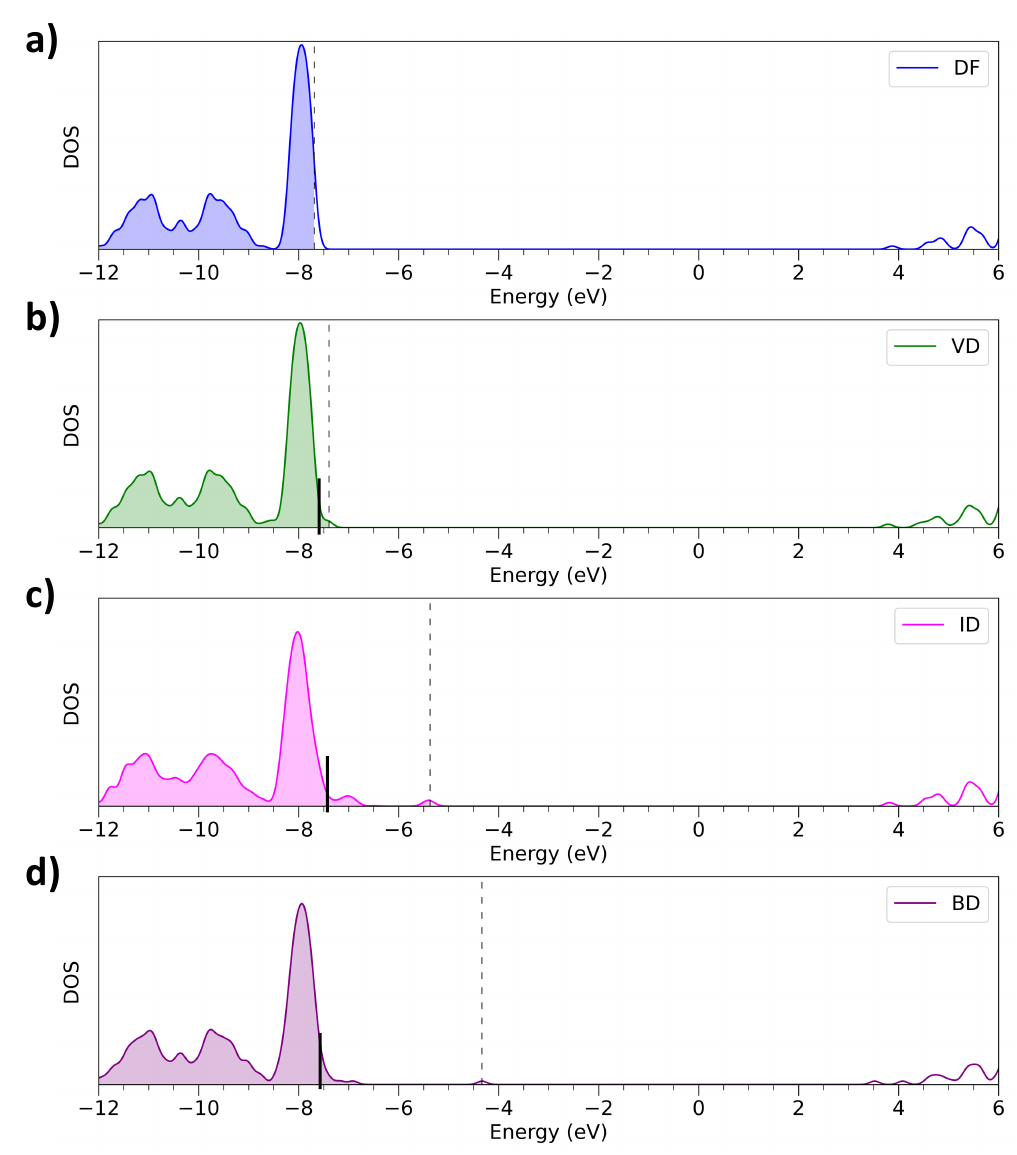}
\caption{Average density of states (DOS) for the defect-free (DF), vacancy (V), ionic (I), and Bjerrum (B) defect configurations are shown in panels a–d as blue, green, magenta, and purple curves, respectively. Occupied states are shaded following the same color scheme, while unoccupied states remain unshaded. The average Fermi level is indicated by a vertical dashed gray line. For each defective system (VD, ID, BD), the corresponding valence band maximum is marked by a vertical solid black line.}
\label{SI_7}
\end{figure}

\begin{figure}
\centering
\includegraphics[width=\linewidth]{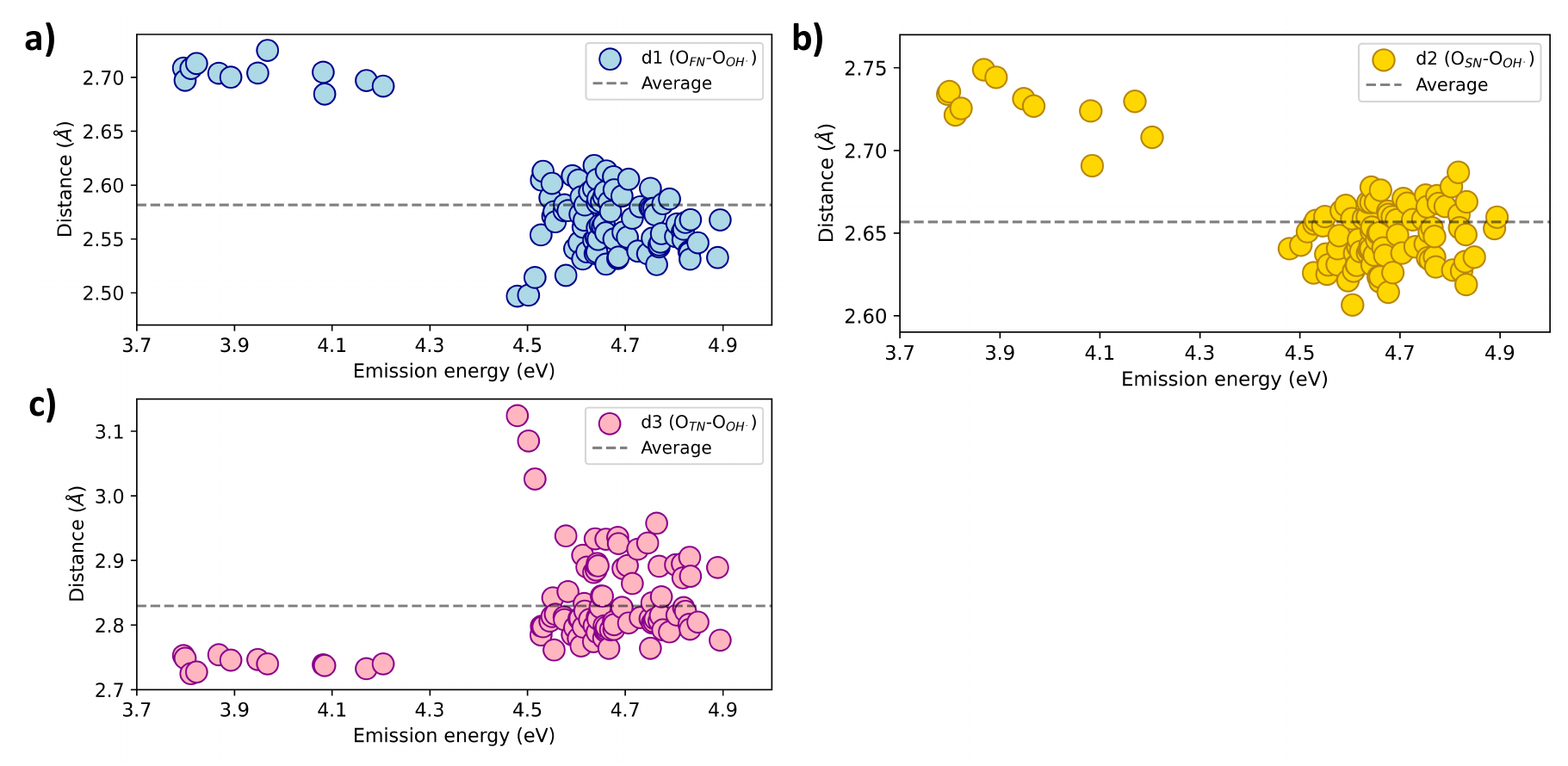}
\caption{Distances between the oxygen atom of the OH$^.$ ion and its first, second, and third oxygen neighbors are shown as a function of the emission energy for 100 vacancy point defect configurations. Panel a) (light blue dots) presents the distances between the oxygen of OH$^.$ and its first nearest neighbor (d1). Panel b) (gold dots) displays the distances between the oxygen of OH$^.$ and its second nearest neighbor (d2). Panel c) (pink dots) shows the distances between the oxygen of OH$^.$ and its third nearest neighbor (d3). The labels FN, SN and TN in the figure correspond to the first, second and third nearest neighbors.}
\label{SI_8}
\end{figure}

\begin{figure}
\centering
\includegraphics[width=\linewidth]{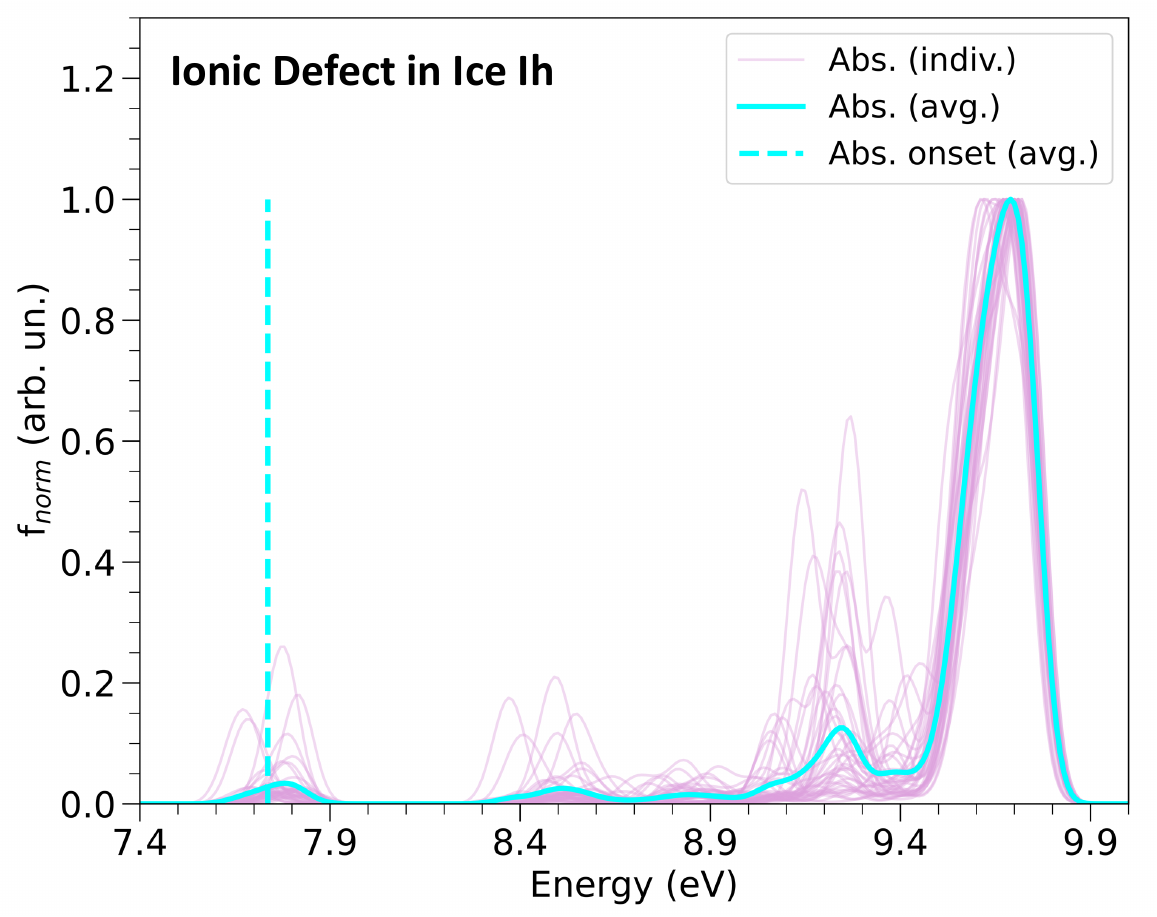}
\caption{Calculated individual (indiv.) optical absorption (Abs.) spectra (solid magenta curves) of 40 ice Ih configurations containing an ionic (OH$^-$) defect. The onset spectrum was obtained by computing the lowest 100 excitations at time-dependent density functional theory level (see the \textit{Methods} section in the main text for details) and applying Gaussian broadening with a $\sigma$ of 0.03 eV and a resolution of 0.01 eV. The corresponding average (avg.) absorption spectrum is shown as a solid cyan curve. The absorption onset value, averaged over the 40 configurations, is reported as a vertical dashed cyan line.} 
\label{SI_9}
\end{figure}

\begin{figure}
\centering
\includegraphics[width=\linewidth]{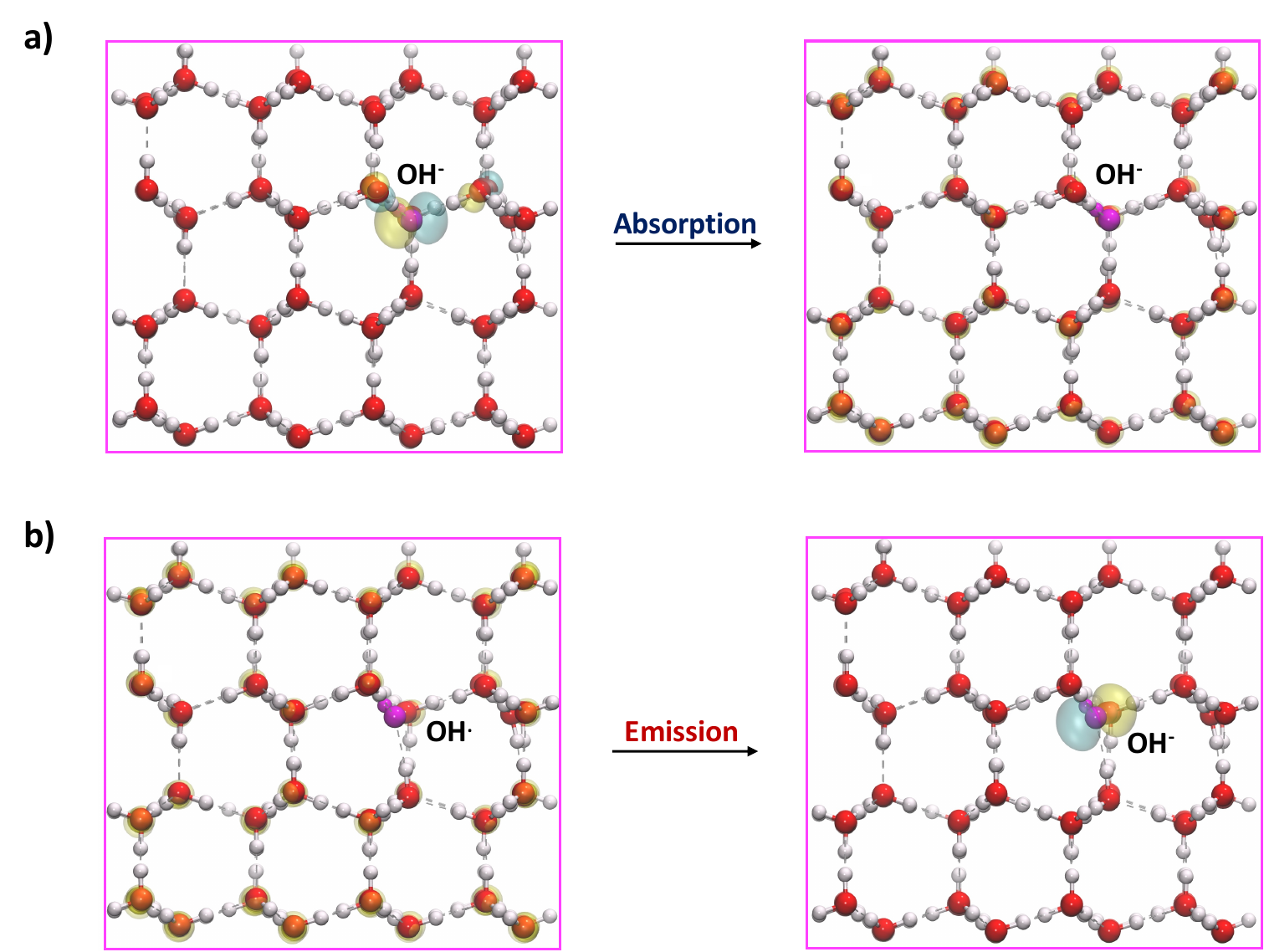}
\caption{The molecular orbitals (MOs) involved in the absorption onset of the model with an ionic (OH$^-$, magenta) defect are shown in panel a. The highest occupied molecular orbital (HOMO) and the lowest unoccupied molecular orbital (LUMO) are displayed on the left and right, respectively. LUMO (left) and HOMO (right) defining the emission onset for the same system are shown in panel b). The hydroxyl radical (OH$^.$), formed in the excited state, is represented as a magenta sphere. In the ground state, the hydroxyl radical recombines with the excess electron to form the hydroxide ion (OH$^-$) which is shown in magenta. In both panels, positive and negative lobes of the MOs are shown in yellow and blue, respectively.}
\label{SI_10}
\end{figure}

\begin{figure}
    \centering
    \includegraphics[width=\linewidth]{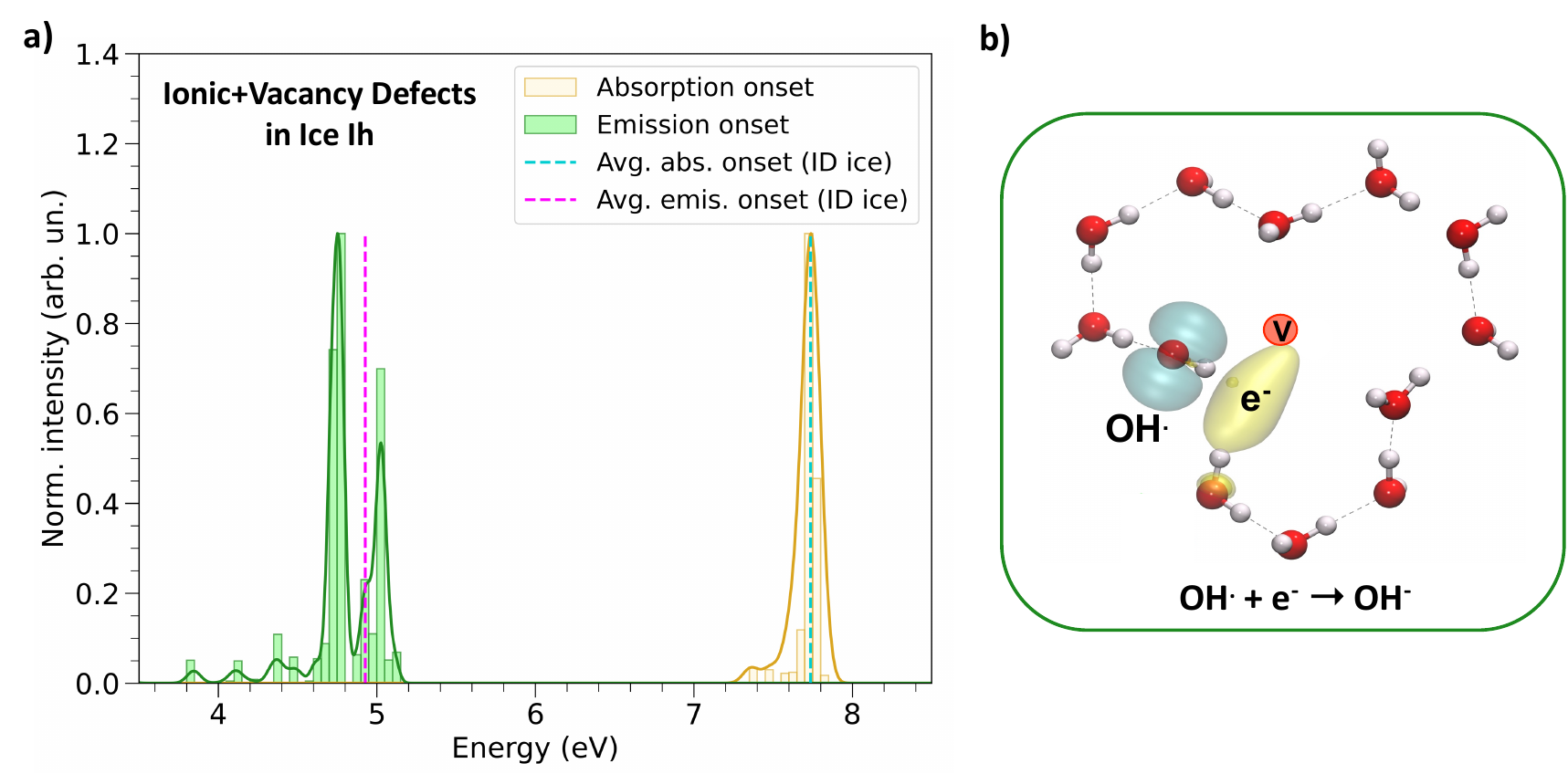}
    \caption{Distributions of absorption (gold) and emission (green) onset energies for 40 configurations of ice Ih containing both an ionic (OH$^-$) and a vacancy defect are shown in panel a. Solid lines represent Gaussian kernel density estimation (KDE) curves fitted to the corresponding normalized histograms. Additional details are given in the \textit{Methods} section in the main text. Average (Avg.) values of the absorption (abs.) and emission (emis.) onset of the ionic defect (ID) ice model are shown as cyan and magenta dashed vertical lines, respectively. An example of unrelaxed differential density is displayed in panel b), showing electron depletion and electron accumulation in blue and yellow, respectively.}
    \label{SI_11}
\end{figure}

\begin{figure}
    \centering
    \includegraphics[width=\linewidth]{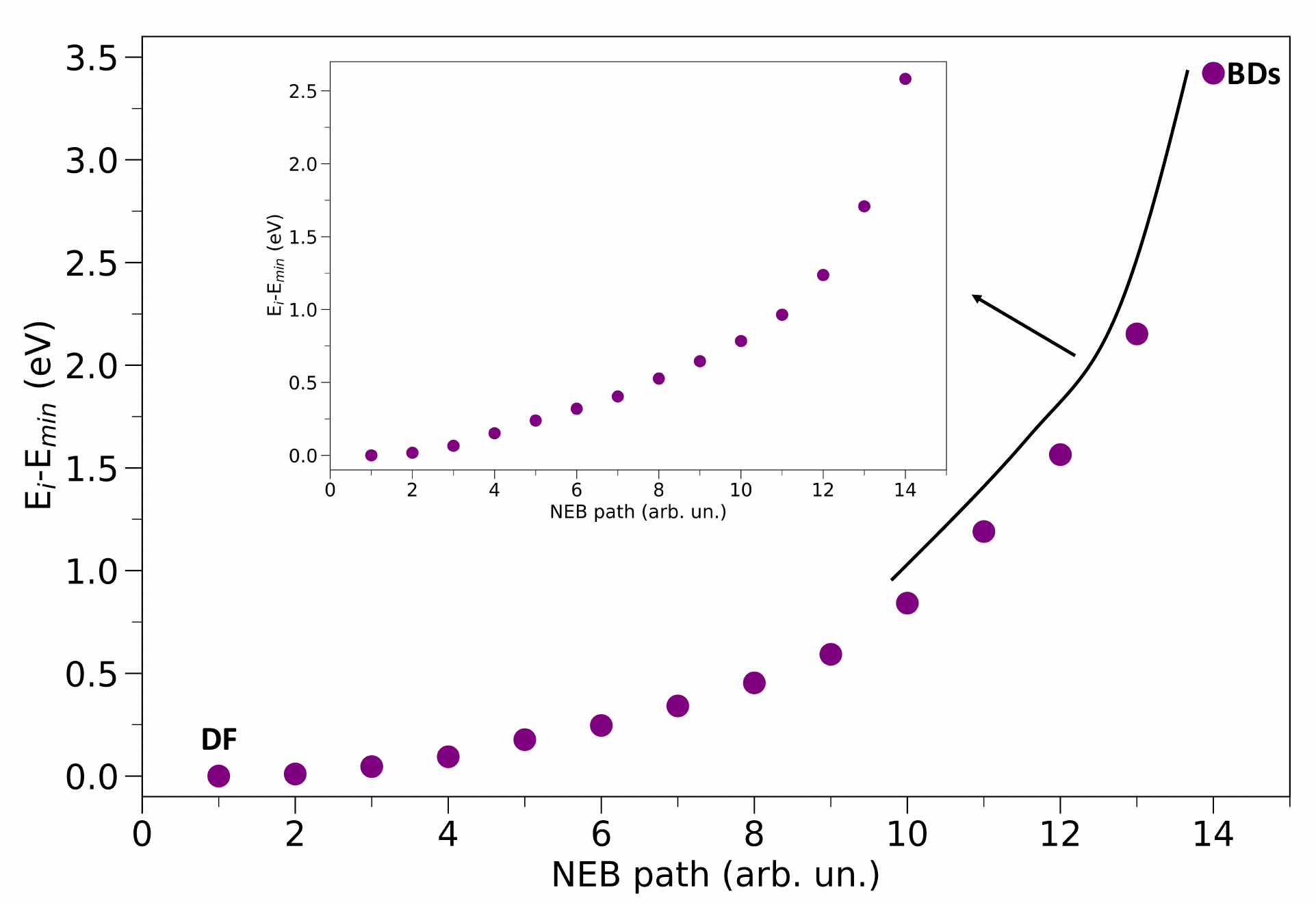}
    \caption{Energy landscape along the nudged elastic band (NEB) pathway connecting a defect-free (DF) configuration to the corresponding excited state optimized Bjerrum defects (BDs) structure.}
    \label{SI_12}
\end{figure}

\begin{figure}
    \centering
    \includegraphics[width=\linewidth]{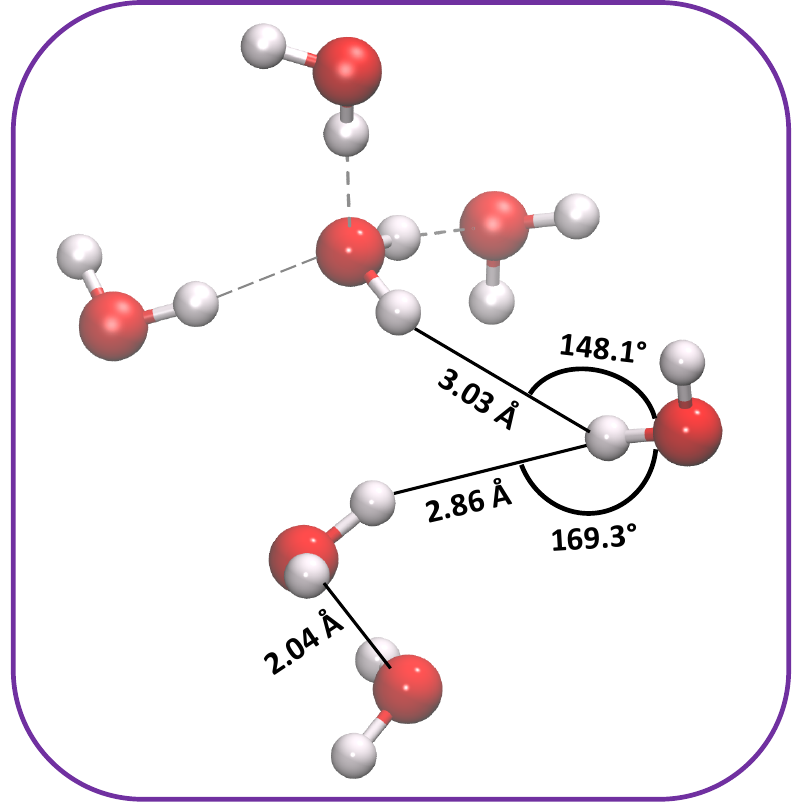}
    \caption{Region of the excited state optimized configuration involved in the formation of the Bjerrum defects. Water molecules comprising the L and D defects, along with an adjacent molecule contributing to an extended D-like defect, are shown as glossy. Surrounding molecules that maintain the hydrogen-bond network with this second D-like molecule are shown as opaque. Reported values include the distances between the two H atoms in each D defect, the corresponding O–H–H angles, and the oxygen–oxygen separation in the L defect, highlighting the local geometry and proton-sharing features that stabilize the defect structure.}
    \label{SI_13}
\end{figure}

\begin{figure}
    \centering
    \includegraphics[width=\linewidth]{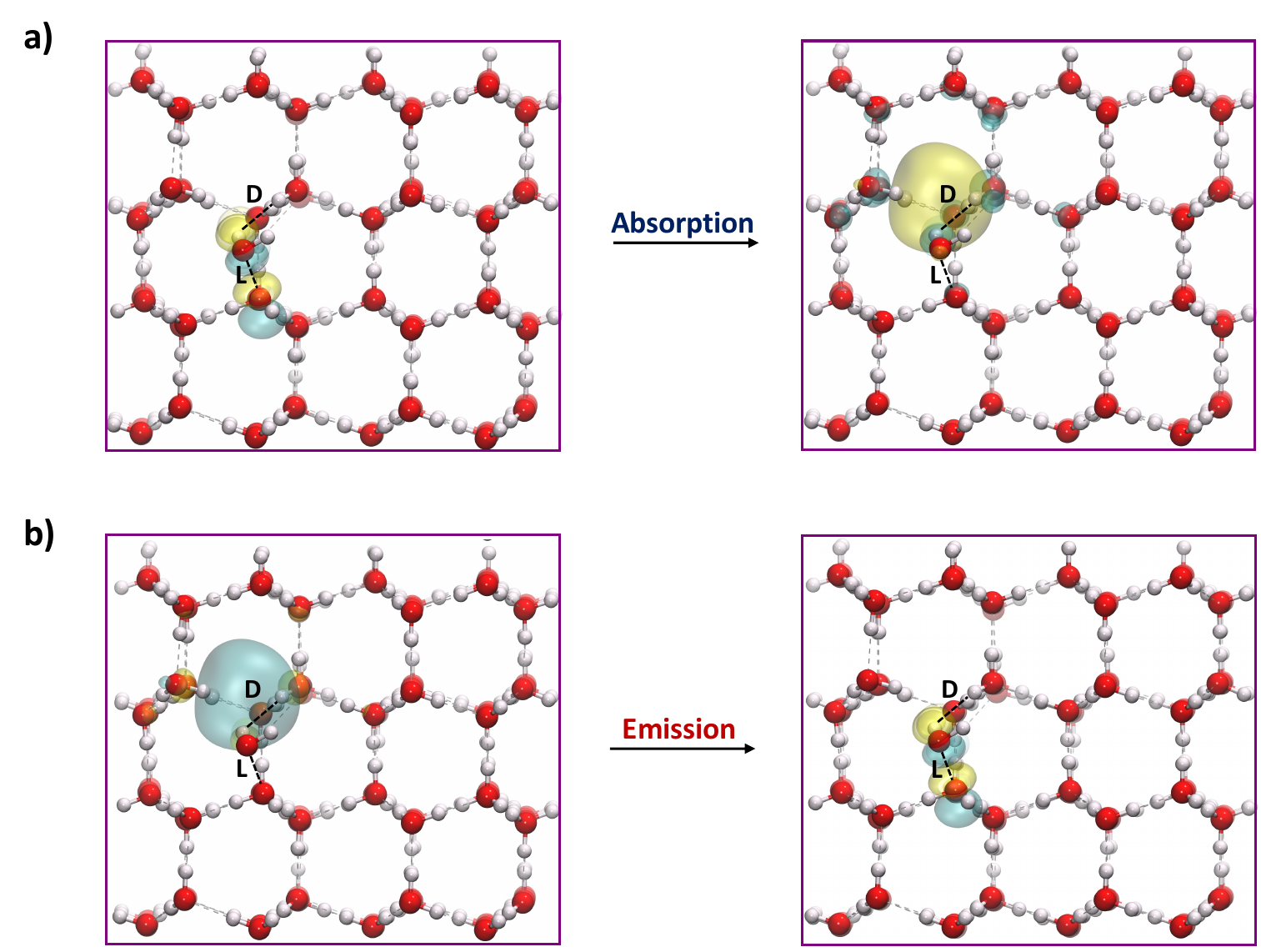}
    \caption{The molecular orbitals (MOs) involved in the absorption onset of the model with D-, and L-type Bjerrum defects are shown in panel a. The highest occupied molecular orbital (HOMO) and the lowest unoccupied molecular orbital (LUMO) are displayed on the left and right, respectively. LUMO (left) and HOMO (right) defining the emission onset for the same system are shown in panel b. In both panels, positive and negative lobes of the MOs are shown in yellow and blue, respectively.}
    \label{SI_14}
\end{figure}

\begin{figure}
    \centering
    \includegraphics[width=\linewidth]{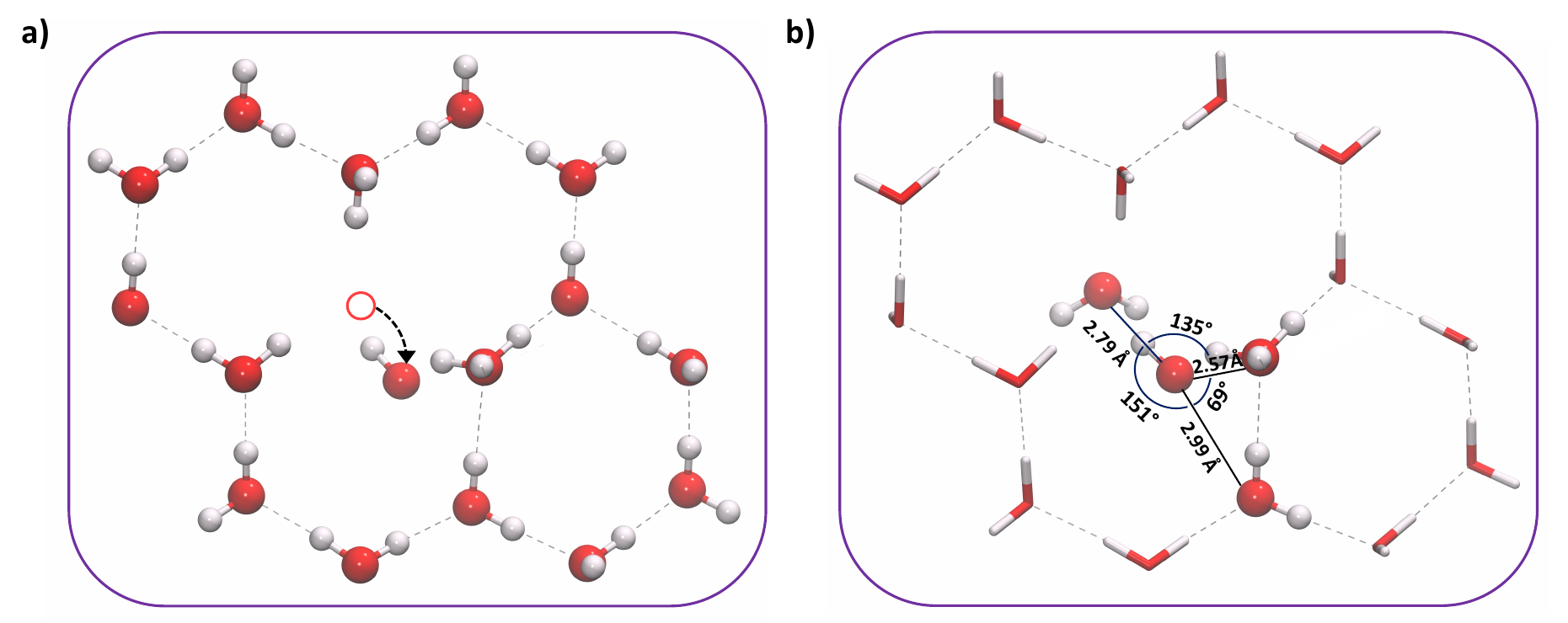}
    \caption{Structure of the defect-free(DF)-like nudge elastic band intermediate, optimized in the excited state, associated with an emission energy of 4.2 eV (see Fig. \ref{Fig5} in the main text). Panel a) highlights the distortion of the hexagonal pattern around the OH$^.$ radical. Panel b) reports the distances and angles between the oxygen of OH$^.$ and its first three nearest oxygen neighbors.}
    \label{SI_15}
\end{figure}

\clearpage

\begin{figure}
    \centering
    \includegraphics[width=0.7\linewidth]{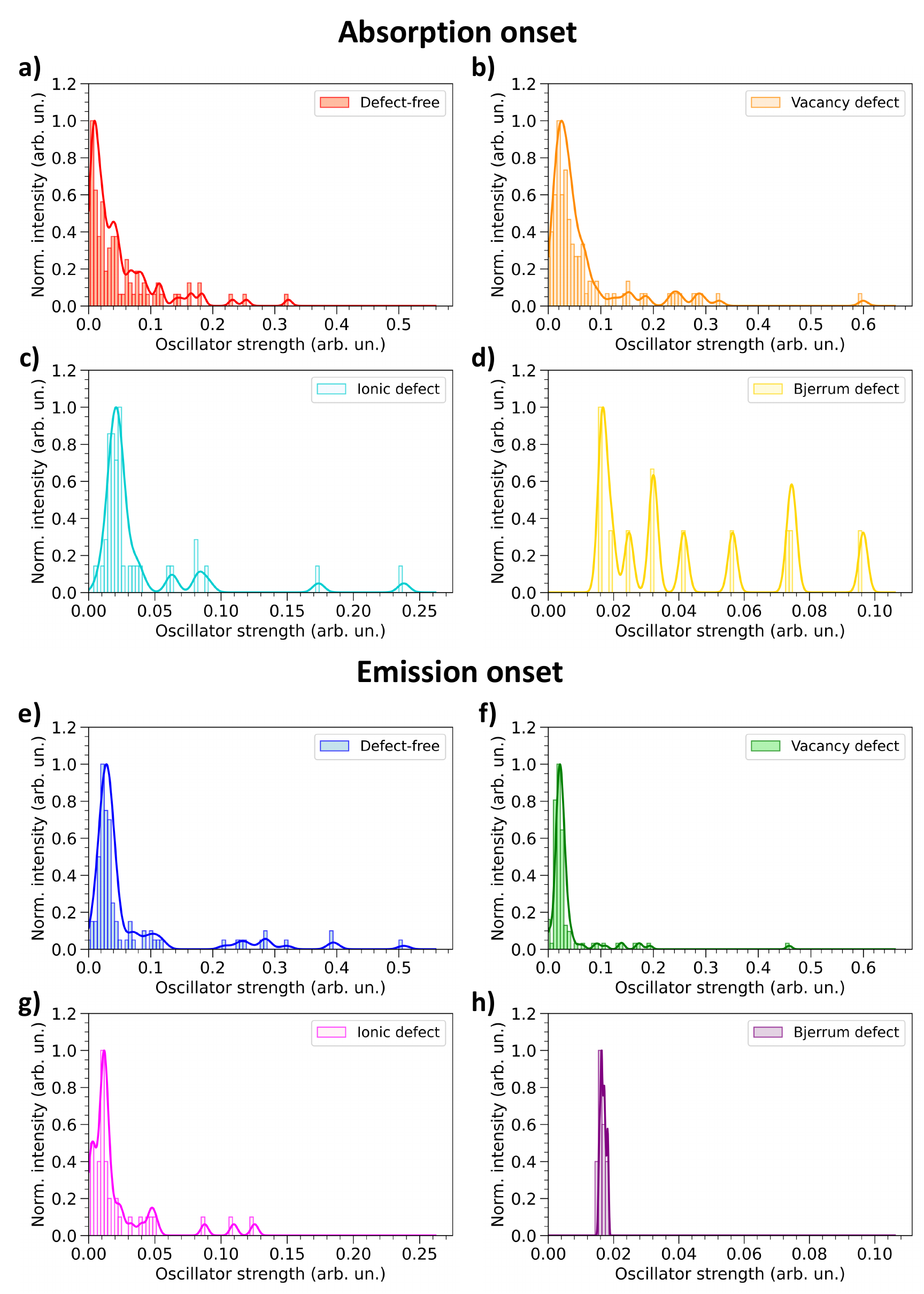}
    \caption{Distributions of absorption (panels a–d) and emission (panels e–h) onset oscillator strength values for defect-free, vacancy, ionic, and Bjerrum defect models. A total of 100 configurations were analyzed for both the defect-free and vacancy models, 40 for the ionic defect model, while 12 Bjerrum-like configurations were obtained from nudged elastic band (NEB) calculations (see Fig. \ref{Fig5} in the \textit{Main Text}). Solid lines represent Gaussian kernel density estimation (KDE) curves fitted to the corresponding normalized histograms. For absorption, the distributions are shown in red (defect-free, a), orange (vacancy, b), cyan (ionic, c), and yellow (Bjerrum, d). For emission, they are shown in blue (defect-free, e), green (vacancy, f), magenta (ionic, g), and purple (Bjerrum, h). Note that for the emission spectra defect-free refers to the initial conditions in the ground state. Of course, the emission spectra for this involve defects as well, such as the hydronium ion, hydroxyl radical, and excess electron.}
    \label{SI_16}
\end{figure}

\begin{figure}
    \centering
    \includegraphics[width=\linewidth]{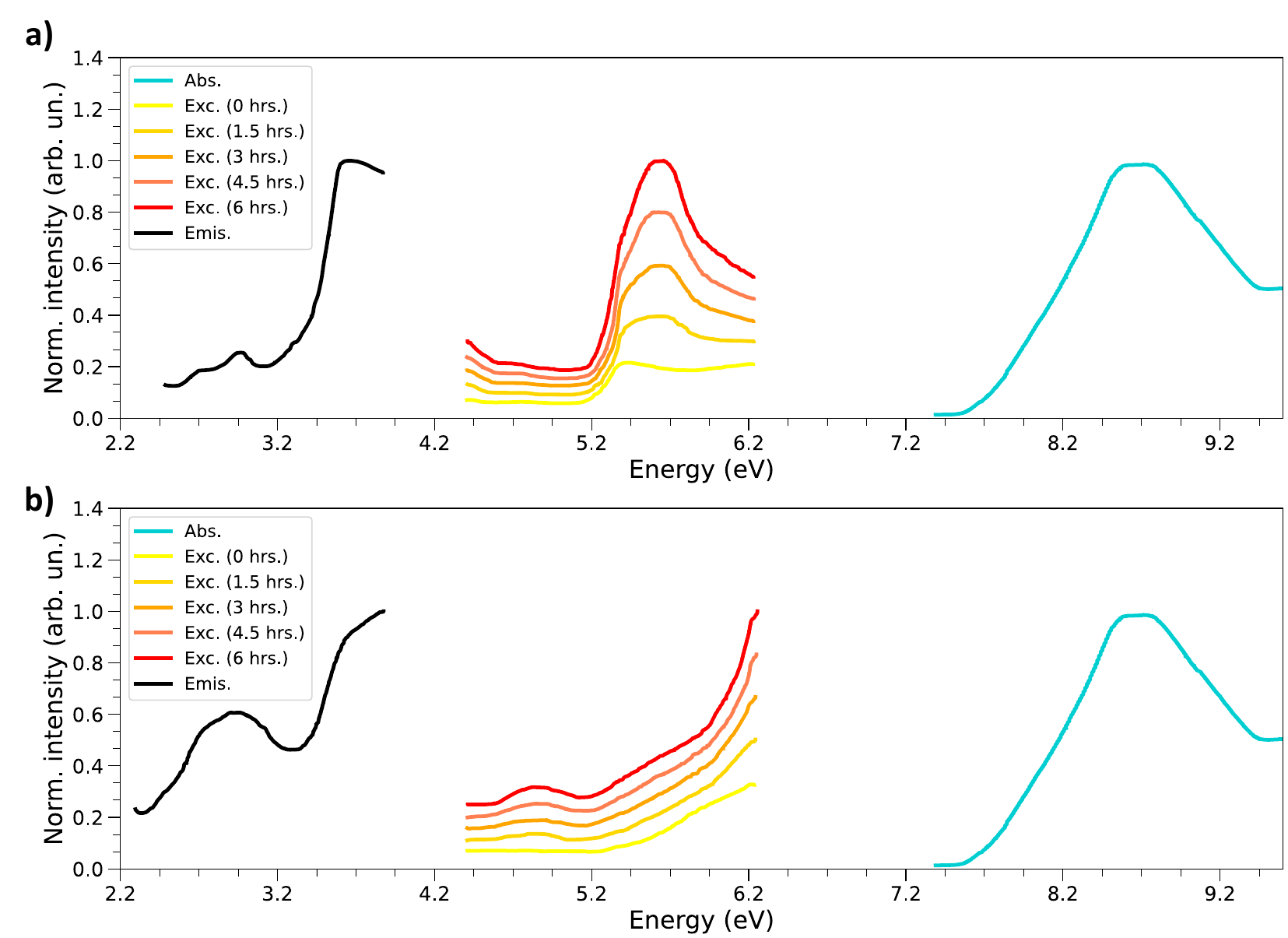}
    \caption{Experimental absorption, excitation, and emission spectra adapted from Refs. \cite{quickenden1985uv,langford2000,kobayashi1983optical}. Excitation spectra (solid yellow, orange, and red lines) were recorded every 1.5 hours over a total of 6 hours under continuous irradiation at $\sim$4.8 eV. The emission wavelength was fixed at 3.65 eV in panel a, and 2.95 eV in panel b. Corresponding luminescence emission spectra (solid black lines), recorded using excitation energies of 5.64 eV (panel a) and 4.77 eV (panel b), are also shown. For reference, the UV absorption spectrum of crystalline ice Ih (solid cyan line), measured in Ref. \cite{kobayashi1983optical}, is included in both panels.}
    \label{SI_17}
\end{figure}

\end{suppinfo}
\clearpage
\bibliography{refs}

\end{document}